\documentclass[a4paper,11pt]{article}
\usepackage{jcappub}
\usepackage[latin1]{inputenc}
\usepackage{amsmath}
\usepackage{amsfonts}
\usepackage{amssymb}
\usepackage{graphicx}
\usepackage{dcolumn}
\usepackage{caption}
\usepackage{tabularx,tikz}
\usepackage{float}
\usepackage{bm}
\usepackage{bbold}
\usepackage{latexsym}
\usepackage{epsfig}
\usepackage[normalem]{ulem}
\usepackage{subcaption}
\usepackage[colorlinks=true]{hyperref}
\usepackage{multirow}
\usepackage{array}

\newcommand{\SM}[1]{\textcolor{red}{\textbf{#1}}}

\usepackage{geometry}
\graphicspath{ {figures/} }

\title{Making maps of cosmological parameters}
\author[a,b,c] {Suvodip Mukherjee,}\emailAdd{smukherjee@flatironinstitute.org}
\author[a,b,c,d]{and Benjamin D. Wandelt}\emailAdd{wandelt@iap.fr}
\affiliation[a]{Institut d'Astrophysique de Paris (IAP), UMR 7095, CNRS/UPMC Universit\'e Paris 6, Sorbonne Universit\'es, 98 bis boulevard Arago, F-75014 Paris, France}
\affiliation[b]{ Institut Lagrange de Paris (ILP), Sorbonne Universit\'es, 98 bis Boulevard Arago, 75014 Paris, France}
\affiliation[c]{Center for Computational Astrophysics, Flatiron Institute, 162 5th Avenue, 10010, New York, NY, USA}
\affiliation[d]{Departments of Physics and Astronomy, University of Illinois at Urbana-Champaign, 1002 W Green St, Urbana, IL 61801, USA}
\date{\today}
\begin{document}
\abstract{ We provide a  fast algorithm to diagnose any directional dependence in the cosmological parameters  by calculating maps of  local cosmological parameter estimates and their joint errors. The technique implements a fast quadratic estimator technique based on Wiener filtering and convolution of the sky with a patch shape. It uses only three map-resolution spherical harmonic transforms per parameter and applies to any data set with full sky or a partial sky coverage. We apply this method to Planck SMICA-2015 and obtain fluctuation map for six cosmological parameters. Our estimate shows that the Planck data is consistent with a single global value of the cosmological parameters and is not influenced by any severe local contaminations. This method is applicable also to other angular or 3D data sets of future missions to scrutinize any local variation in the cosmological parameters.
}
\maketitle{} 
\section{Introduction}
The successful history of several cosmological missions in CMB  \cite{1965ApJ...142..419P, firas, firasan, fm2002, cobe, Bennett:2012zja, Adam:2015rua, Hanson:2013hsb, actpol, Array:2016afx, Ade:2013gez, boomerang, Stompor:2003kr} have paved the path  of precision cosmology and several upcoming missions like EUCLID  \cite{2012sngi.confE...3M}, LSST \cite{2009arXiv0912.0201L}, SKA \cite{Carilli:2004nx}, PIXIE \cite{Kogut:2011xw} and LiteBIRD \cite{Matsumura:2013aja} have the potential to vastly enhance the precision with which we know the cosmological parameters. With increasing signal-to-noise, and more complex inference tasks, systematics will become the primary challenge to cosmological parameter inference. The accurate estimation of cosmological parameters from these next generation missions require advanced methods and sophisticated techniques to accurately remove systematic errors and non-cosmological contaminations from the observed data. 

Inferring parameters from a cosmological data set requires a great deal of care. The approach is typically Bayesian where selection of robust features of the data, detailed statistical modelling of the signal, the systematics, and the measurement process leads to a likelihood  that is combined with carefully chosen priors. The resulting posterior probability density for the parameters is then usually explored by sampling it with Markov Chain Monte Carlo (MCMC) \cite{Lewis:2002ah}. Standard analyses assume statistical homogeneity and isotropy of the universe, a natural assumption  since there is no strong evidence for violations of these symmetries. However,  the presence of systematics like foreground contamination, anisotropic noise and beam, and other perhaps incompletely modelled instrumental, environmental or astrophysical systematics, may introduce biases in the parameters.  These systematics would not be expected to respect the rotational or translational symmetries of an isotropic and homogenous universe. So a diagnostic that estimates parameters on many sky patches and visualises the way these estimates vary across the sky, that is a \textit{parameter map}, would  be a useful tool to test for the presence of systematics and to obtain clues as to their origin.  

In addition, in the absence of strong systematics such an analysis has the potential to reveal a first hint of violations of the isotropy or homogeneity assumptions. Even in a globally symmetric universe, our sky could contain statistical anisotropy of cosmological origin. For examples, physical parameters could take different values in different domains and we might sit between two or more such domains. This approach was taken in \cite{Javanmardi:2015sfa, Carvalho:2015lqd, Carvalho:2016aah} to test for direction dependence of the cosmological parameters from Supernovae data. Any such hints could be followed up with dedicated searches, testing specific physical models that would predict such variations \cite{Cai:2011xs, 2013ApJ...773L...3A, 2016JCAP...06..042M, 2015JCAP...11..012H, 2001PThPh.105..419T, McClure2007533, Romano:2014iea, PhysRevD.93.083525, Mukherjee:2015wra}. The presence of enhanced temperature fluctuations in the local patches of the sky is also of interest due to  predictions from theoretically motivated models and were studied in past using WMAP data \cite{1998CQGra..15.2657C, 2011ApJ...740...52H, 2010ApJ...724..374K, 2011JCAP...04..033M, 2013MNRAS.429.1376B, 2010JCAP...02..004F, 2011PhRvD..84d3507F}.

MCMC samplers are now the standard approach to cosmological parameter inference. But to run an MCMC chain in  each patch to infer the cosmological parameters is expensive. In this paper, we devise a  fast, efficient and simple algorithm to estimate  directional dependence of cosmological parameters and implement it on  CMB maps. We show that the cosmological parameters can be computed through a quadratic estimator that implicitly projects a local deviation of the angular power spectrum of the signal onto a parameter change from a fiducial value. The appropriate projection is done by applying a an optimal  filter. Optimal or Wiener filtering is a  powerful method that is widely used in  cosmology data analysis in several contexts \cite{Tegmark:1996qt, Bond:1998zw, Oh:1998sr, Wandelt:2003uk, Elsner:2012kj, Hinshaw:2003fc, Dunkley:2008ie, Elsner:2012bq, Tegmark:1996qs, Hirata:2004rp, Smith:2007rg, Elsner:2012fe,Bunn:2016lxi, Ramanah:2017luo}. 

The variation in the angular power spectrum of the CMB temperature field can be related to leading order to the variation in cosmological parameters by performing a Taylor series expansion. Our method shares this feature with the method described in \cite{2002MNRAS.334..167G} but goes beyond it by showing how to do parameter analysis on many patches without having to compute the power spectra  explicitly for the cost of a very mild approximation. The result is a map of parameter estimates and correlated uncertainties. Owing to its computational simplicity this method can be implemented on large data sets even with complicated sky coverage efficiently and with limited computational cost. 

The paper is organized as follows. In Sec.~\ref{formalism}, we lay out the basic formalism of this method. In Sec.~\ref{procedure}, we discuss the implementation of this method on CMB temperature maps. We obtain the map for variation of cosmological parameters from the publicly available Planck SMICA map \cite{smica}  and compare it with  a simulated CMB map in Sec.~\ref{results}.
 We summarize and conclude in Sec.~\ref{conc}.


\section{How to map cosmological parameter variations}\label{formalism}
\subsection{Formalism for arbitrary parameter patches}
To achieve our goal of developing a fast diagnostic to explore the data we will want to perform parameter analyses on many patches of the data. Each patch will necessarily  provide a weaker constraint on the parameters than the full data set.  As a consequence we will content ourselves with an asymptotic approach, that returns a parameter estimate for each patch, together with an asymptotic error estimate, as follows. 

We develop the method for the example of the cosmic microwave background temperature (CMB) fluctuations on the sphere but applies without change to any other scalar field on the sphere. Fields on rectangular domains can be treated strictly analogously by replacing spherical harmonics with Fourier coefficients.  The discussion can also be generalized to spin-$n$ fields, such as CMB polarization anisotropies or lensing shear measurements without difficulty.

The CMB data map containing the measured temperature and polarization anisotropies  $d(\hat n)$ can be expressed in the spherical harmonic basis as
\begin{equation}\label{eq1}
d(\hat n) = \sum d_{lm}Y_{lm}(\hat n),
\end{equation}
where $d_{lm}$ have zero mean, are mutually uncorrelated and have variance $C_{l}$, the power spectrum.
For full sky, noise-free data the optimal estimate of the $C_{l}$ is
\begin{equation}\label{eq2}
\hat C_l= \frac{\sum_{m}d_{lm}d^*_{lm}}{2l+1}, \quad l=2,\dots , L,
\end{equation}
which can be written using  $\bm{P}_{l}$, the operator projecting onto  angular wavenumber $l$
\begin{equation}
\hat C_{l}=\frac1{2l+1}\bm{d}^{T}\bm{P}_{l}\bm{d}.
\end{equation}

For the estimation of the cosmological parameters in different patches of the sky, each patch will be defined by the patch shape window $W^{p}(\hat n_j) $,
\begin{equation}\label{eq:patchwindow}
W^{p}(\hat n_j) = \begin{cases}& 1 \,\, \text{if pixel $j$ is in  patch $i$},\\
&0\,\, \text{otherwise}\end{cases}
\end{equation}
for the case of a sharp-edged, unapodized patch window. It will be convenient to define the operator $\bm{W}^{p}$ as the diagonal matrix with elements $W^{p}_{jj}=W^{p}(\hat n_j)$  such that the map of temperature fluctuations in patch $i$ is $\bm{W}^{p}\bm{d}$. The pseudo power spectrum  on this patch $\tilde C_l ^{p}$  \cite{1973ApJ...185..757H, Wandelt:2000av} is defined as 
\begin{equation}\label{cl-0}
\tilde C_l ^{p}=\frac1{2l+1}(\bm{W}^{p}\bm{d})^{T}\bm{P}_{l}\bm{W}^{p}\bm{d},
\end{equation}
provides a statistical summary of the information the patch $i$ contains about the cosmological parameters\footnote{In fact, any power spectrum estimate may be used. In the faster, harmonic-space variant of our method described section~\ref{sec:symmetricPatches} we specialize to a pseudo-spectrum estimator  to achieve a drastic reduction in computational cost. The well-known pseudo-spectrum is a member of this class.}. In expectation, this is related to the full sky  power spectrum $C_{l}$ by the mode coupling matrix $M_{ll'}  ^{p}$ imposed by the patch shape and the noise bias $N_{l}^{p}$ 
\begin{equation}\label{cl-1}
\langle \tilde C_l ^{p}\rangle= \sum_{ll'} M_{ll'}^{p} \langle C_{l'}\rangle +N_{l}^{p}.
\end{equation}
We have omitted the instrumental transfer function and the beam for clarity of presentation; those can be included in the usual way, as shown in  \cite{Hivon:2001jp}. 

For a fiducial set of $n_{\theta}$ cosmological parameters $\theta^{\text{fid}}_{j}$, $j=1,\dots, n_{\theta}$, let the predicted global CMB power spectrum be $ C^{\text{fid}}_{l}\equiv C_{l}(\theta^{\text{fid}})$.
As the cosmological parameters determine the power spectrum of the CMB, any patch-dependence of the cosmological parameters translates into a patch-dependence of the power spectrum.  Defining $\theta_j^{p} $ to be the cosmological parameters in  patch $p$, 
 the deviation in the value of the measured spectrum $\tilde C_l ^{p}$ from  that expected in the fiducial model can be written as
\begin{align}\label{cl-2}
\begin{split}
\delta \tilde C_l ^{p}= \tilde C_{l} ^{p}- \tilde C^{\text{fid}\;p}_l -N_{l}^{p}, \text{where} \,\, \tilde C^{\text{fid}\;p}_l= \sum_{ll'} M_{ll'} ^{p}C^{\text{fid}}_{l'}.	
\end{split}
\end{align}
The  $\delta \tilde C_l ^{p}$ capture any deviation of the power spectrum in patch $p$ from that which is expected given the fiducial, global model and the mode-coupling induced by the patch shape.

 Writing up to linear order in the parameter variation, we have  
\begin{equation}\label{cl-4}
\delta \tilde C_l ^{p}= \sum_j^{n_{\theta}}\frac{\partial \tilde C^{\text{fid}\;{p}}_l}{\partial \theta_j}\bigg|_{\theta_j=\theta^{\text{fid}}_j} (\theta_j ^{p} - \theta^{\text{fid}}_j) +\dots .
\end{equation}

In the following, we switch to matrix notation. We write an $(L-1)$-vector $\delta\bm{c} ^{p}$ for the quantity in Eq.~\ref{cl-2}  and collect the derivatives into the $(L-1)\times n_{\theta}$-matrix $\bm{D}^{p}$.   The parameters are the $n_{\theta}$-vector $\bm{\theta}^{p}$. Then Eq.~(\ref{cl-4}) becomes
\begin{align}\label{cl-5}
\begin{split}
\delta\bm{ c} ^{p}= \bm{D}\left(\bm{\theta}^{p}-\bm{\theta}^{\text{fid}} \right)=
 \bm{D}^{p}\delta\bm{ \theta} ^{p},
\end{split}
\end{align}
where the last equality defines $\delta \bm{\theta} ^{p}$ which captures the variation from fiducial values of the  cosmological parameters in the patch of the sky with the center of the direction given by $\hat n_i$. 

The measurement covariance matrix  for the power spectrum in a given patch $\hat n_i$ is $\bm{K}_{\delta \bm{c}}^{p}$, with components $ K_{\delta c\;ll'}= \langle (\delta \tilde C_l)(\delta \tilde C_{l'})^t \rangle$. With these ingredients we can now  compute an asymptotically optimal estimator for the $\delta \bm{\theta}$  in terms of  the $\delta \bm{c} $ in each patch as follows.

Working in the asymptotic regime where the likelihood for $\bm{c}$ is Gaussian to a good approximation,  the maximum likelihood estimate for the parameter deviation in each patch is
\begin{equation}\label{eq:generalestimator}
\hat{\delta\bm{\theta}}^{p} = \left(\bm{F}^{p}\right)^{-1}\bm{D}^{p\;T} \left(\bm{K}_{\delta \bm{c}}^{p}\right)^{-1} \delta \bm{c}^{p},
\end{equation}
with covariance 
\begin{equation}
\bm{K_{\theta}}^{p}=\left\langle\delta \hat{\bm{\theta}} ^{p}
\delta \hat{\bm{\theta}}^{p\;T}\right\rangle=\left(\bm{F}^{p}\right)^{-1},
\end{equation}
where $\bm{F}^{p}$ is the Fisher matrix in  patch $p$ (see \cite{Tegmark:1996bz} for a discussion of the Fisher matrix).  In the asymptotic case we are considering here 
\begin{equation}\label{fisher-def}
F_{ij}=\bm{D}^T\bm{K}_{\delta \bm{c}}^{-1}\bm{D}.
\end{equation}

 Note that given the data vector $\bm{c}$, the parameter estimate $\hat{\bm{\theta}} =\bm{\theta}^{\text{fid}}+\delta \hat{\bm{\theta}}$ saturates the Cramer Rao bound;  the parameter estimates are optimal.  This result is analogous to  the result of \cite{2002MNRAS.334..167G} which focuses on compressing the data to derive a  Fisher-optimal likelihood for MCMC exploration. Our estimate comes from the direct maximization (without running an MCMC chain) of the same, optimal likelihood for every patch.

The method as described so far applies under quite general circumstances such as when the individual patches have very different shapes or noise properties. Since the method does not require running MCMC chains it is fast enough to run an analysis on 48 sky patches from the Planck data in about one hour on a standard laptop. The parameter estimates in different patches can be computed efficiently in parallel. We will discuss additional computational considerations in subsection~\ref{sec:computationalcost}.

We will now turn to a harmonic space technique that accelerates computing the parameter maps under some additional assumptions.
 

\subsection{Azimuthally symmetric patches}\label{sec:symmetricPatches} 
Assume that we choose the patches as  (possibly apodized) circular disks  centered on all pixels of a parameter map\footnote{Any other azimuthally symmetric patch shapes, such as annuli can be treated without modification of the formalism.}. All patches have identical size, shape, and radial profile. The patch window is then a function of the angular distance to the patch center $\hat n_p$ only,
\begin{equation}\label{mag-2}
W^{p}(\hat n_i) \equiv W(\hat n_p.\hat n_i) = \sum_{l} \frac{2l+1}{4\pi}  W_{l} P_{l}(\hat n_{p}.\hat n_{i}).
\end{equation}
As a consequence, the coupling matrix $M_{l l'}$, the $\hat C^{\text{fid}}_l$, and the $\bm{D}$, are the same for all patches. If the beam, instrumental  transfer function and power spectrum co-variance can also be approximated as the same for all patches we can use the following harmonic-space method.\footnote{These assumptions will break down for patches that intersect the survey mask. For quick analysis, these edge cases might simply be ignored. If desired, they can be treated separately by the method described in the previous subsection. Pixels masked to exclude bright point sources will occur all over the map, but the coupling matrix $M_{ll'}$ only depends on the average number of these exclusions per steradian and may therefore be treated as  constant across the sky to a good approximation. If the noise power varies significantly across the sky, the resulting noise bias in the parameters can be corrected for simply by subtracting the effect on parameters of the noise bias term in Eq.~\ref{cl-1} by computing the average  parameter map on  signal-free noise simulations. We ignore the variation this would induce for the covariances of the parameter estimates, which could again be estimated with a modest number of Monte Carlo simulations.} We will drop the explicit patch superscript for quantities that are the same for all patches.

Let us compute the parameters for one such circular patch. From Eq.~\ref{eq:generalestimator} we find that 
\begin{equation}\label{eq:estimator}
\delta\hat{\bm{\theta}} ^{p}= \bm{F}^{-1}\bm{D}^T \bm{K}_{\delta \bm{c}}^{-1} \delta \bm{c}^{p},
\end{equation}

We can define the rectangular $(n_{\theta}\times L)$-matrix  with elements
\begin{equation}
{O}^{j}_{l}\equiv \sum_{k,l'}{F}^{-1}_{jk}{D}^{T}_{kl'} {K}_{\delta \bm{c}\,{l'l} }^{-1}\frac1{2l+1},
\end{equation}
Considering each of the $n_{\theta}$ rows separately\SM{,} each element of  $\delta\hat{\bm{\theta}}$  in equation~\ref{eq:generalestimator} can be computed simultaneously for all pixels in the parameter map, as follows.

We want to extract from the map the portion whose fluctuations compress the information about cosmological parameters. The parameter components of equation~\ref{eq:estimator} can be rewritten as 
\begin{align}
\delta{\hat{\theta}}_{j} ^{p}&=\sum_{l}{O}_{jl}\left[(\bm{W}^{p}\bm{d})^{T}\bm{P}^{l}\bm{W}^{p}\bm{d}-({\tilde C}^{\text{fid}}_{l}+N_{l})\right],\\
&=(\bm{W}^{p}\bm{d})^{T}\bm{O}^{j}\bm{W}^{p}\bm{d}-b^{p}_{j},
\label{eq:symmetricpatches}
\end{align}
where we defined $\bm{O}^{j}\equiv \sum_{l}{O}^{j}_{l}\bm{P}^{l} $ and $b^{p}_{j}$ is the  sum of the expected isotropic signal and noise contribution to the parameter estimate in the patch.

Aside from reducing the number of pre-computations due to the simplified setup this expression is no faster to evaluate than Eq.~\ref{eq:estimator}. But approximating this expression by reversing the order of the operators $\bm{O}^{j}$ and $\bm{W}^{p}$ in the first term of~\ref{eq:symmetricpatches} results in 
\begin{align}
(\bm{W}^{p}\bm{d})^{T}\bm{O}^{j}\bm{W}^{p}\bm{d} \approx(\bm{W}^{p}\bm{d})^{T}\bm{W}^{p}\bm{O}^{j}\bm{d}=\sum_{i}W^{p\;2}_{i}d_{i} [\bm{O}^{j}d]_{i},
\label{eq:approx}
\end{align}
the dot product of the patch window with the quadratic map obtained by the pixelwise multiplication of the filtered data map with its unfiltered version. The parameters in \textit{all} patches can therefore be obtained  using the following series of operations: 1) Transform the data map into harmonic space; 2) Generate $n_{\theta}$ maps by multiplying with each $O^{j}$ and transforming back to pixel space; 3) Multiply every pixel of this map by the corresponding pixel in the data map; 4) Then convolve each of the resulting maps with the square of the patch window in harmonic space to produce the parameter maps at the desired resolution. This last step can be done efficiently in harmonic space, and the final transform to pixel space needs only to be done at the coarser patch resolution (rather than the full pixel resolution) to produce the parameter map.

In total this requires $1+3n_{\theta}$ harmonic transforms covering the full bandwidth and resolution of the data map, and then another $n_{\theta}$ low-bandwidth and low-resolution transforms to produce the parameter maps on a coarse pixelisation. The full resolution transforms dominate the scaling in the relevant regime and give $O\left((1+3n_{\theta}) n_{\text{pix}}^{3/2}\right)$. Contrast this with the $n_{\text{patch}}$ full transforms to produce the $\delta \bm{c}^{p}$  and the transforms required to compute the coupling matrices that dominate the $O\left(2n_{\text{patch}}n_{\text{pix}}^{3/2}\right)$ computational cost  Eq.~\ref{eq:estimator}. We will compare the computational aspects of the two methods in more detail in section~\ref{sec:computationalcost}.

\subsection{Nature of the approximation}
The map obtained by using Eq.~\ref{eq:approx} differs slightly from the one defined in Eq.~\ref{eq:symmetricpatches}, even for azimuthally symmetric patches. The difference comes from the way the patch window is applied. By design, only pixels within the patch $p$ affect the parameter estimate for that patch in  Eq.~\ref{eq:symmetricpatches}. The estimates from Eq.~\ref{eq:approx} however result from the application of the optimal filters $\bm{O}^{j}$ to the entire sky map (after masking of foreground contaminated regions) and then measuring the variance within sky patch $p$.

Since the real-space convolution kernels corresponding to the  $\bm{O}^{j}$ have finite support parameter estimates in neighbouring patches will be more strongly correlated than would be the case in the original approach Eq.~\ref{eq:symmetricpatches}; one the other hand the parameter estimates would be less noisy, since acting with the kernels on the full sky would allow a cleaner separation of the contributions each parameter makes to the variance in the patch.

Radially truncating the pixel-space kernels corresponding to the $\bm{O}^{j}$ would improve the approximation to affect only immediately adjacent patches. This can be done simply by Legendre transforming the rows of $O^{j}_{l}$, truncating at the cosine of angular radius of the patch kernel and transforming back. This recovers the same $d_{i} [\bm{O}^{j}d]_{i}$ for pixels in the center of the patch but would still couple the edge of this map to pixels in neighbouring patches. We find the  effect of the approximation to be modest in our numerical tests, and did therefore not implement the kernel truncation in any of the applications presented in this paper.

\subsection{Implementation on CMB simulations}
 
\begin{figure}[H]
    \centering
    \begin{subfigure}[b]{0.45\textwidth}
        \includegraphics[trim={0 0 0 0.8cm},clip,width=\textwidth]{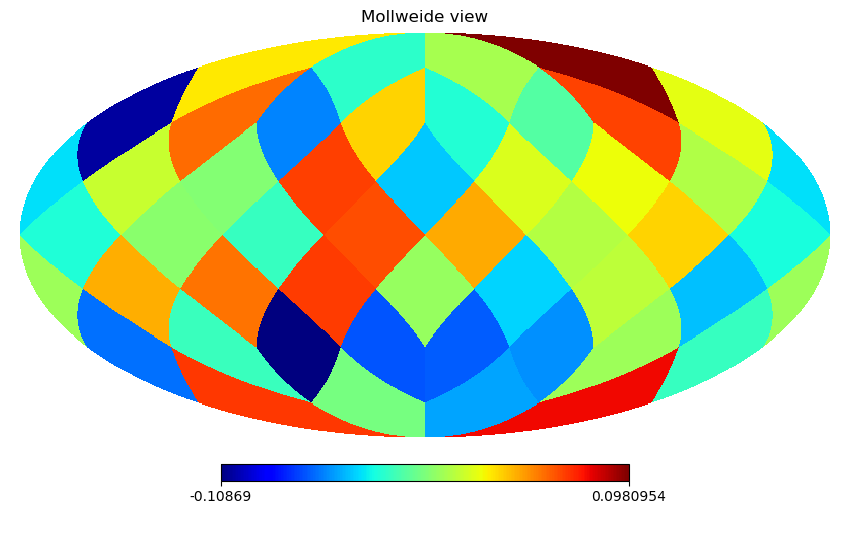}
        \caption{$A_s$}
    \end{subfigure}
    ~ 
    \begin{subfigure}[b]{0.45\textwidth}
    \includegraphics[trim={0 0 0 0.8cm},clip,width=\textwidth]{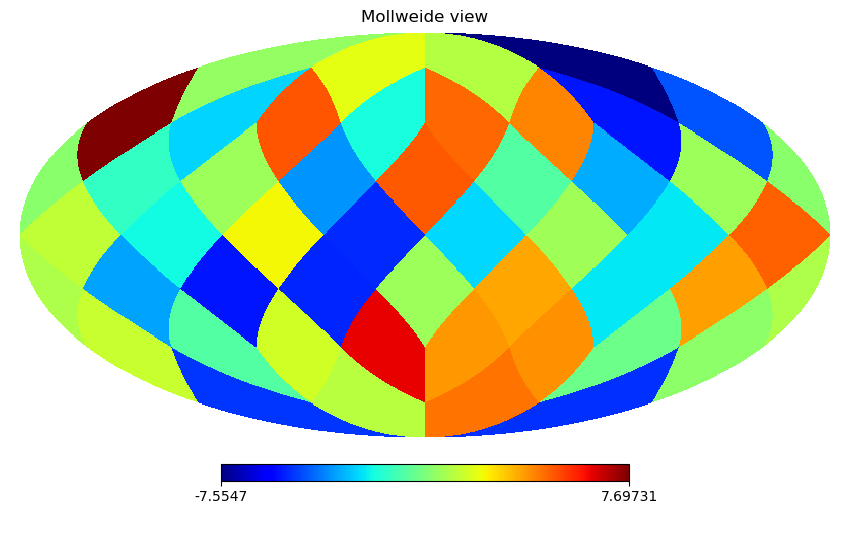}
        \caption{$H_0$}
    \end{subfigure}
    \begin{subfigure}[b]{0.45\textwidth}
   \includegraphics[trim={0 0 0 0.8cm},clip,width=\textwidth]{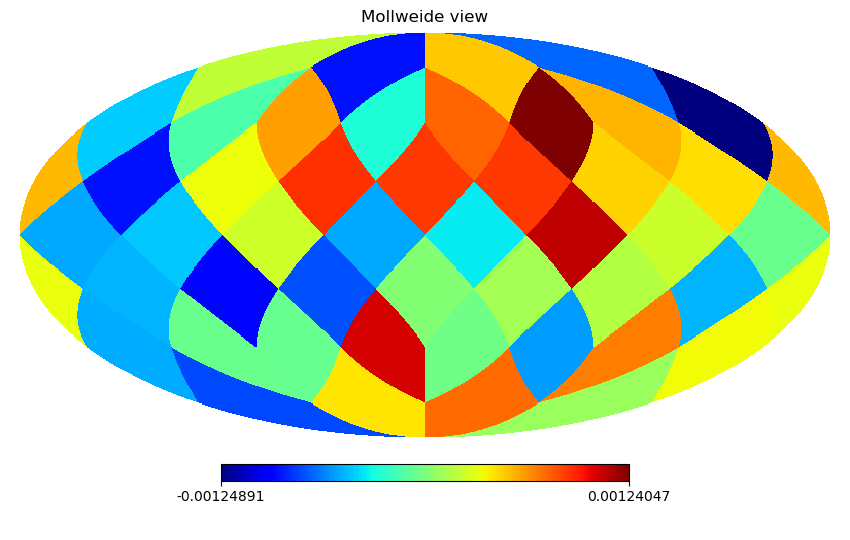}
        \caption{$O_bh^2$}
    \end{subfigure}
    ~ 
    \begin{subfigure}[b]{0.45\textwidth}
        \includegraphics[trim={0 0 0 0.8cm},clip,width=\textwidth]{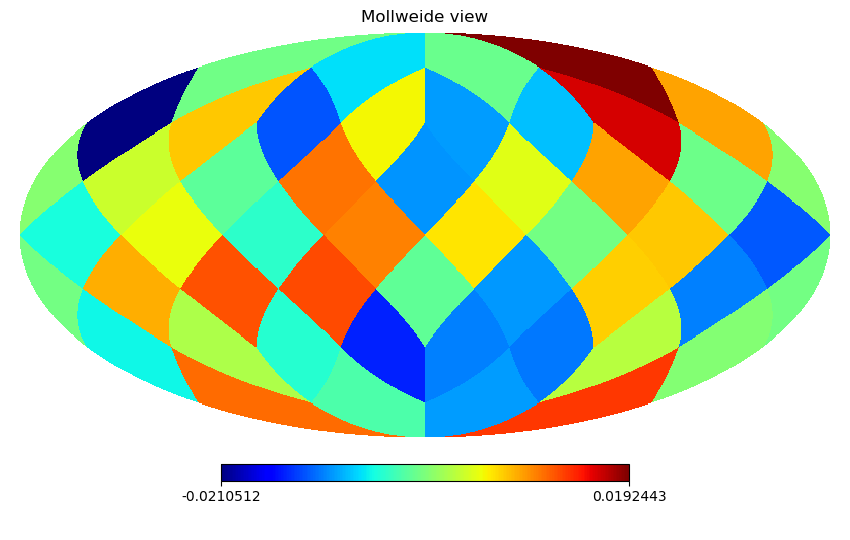}
         \caption{$O_ch^2$}
    \end{subfigure}
    \begin{subfigure}[b]{0.45\textwidth}
    \includegraphics[trim={0 0 0 0.8cm},clip,width=\textwidth]{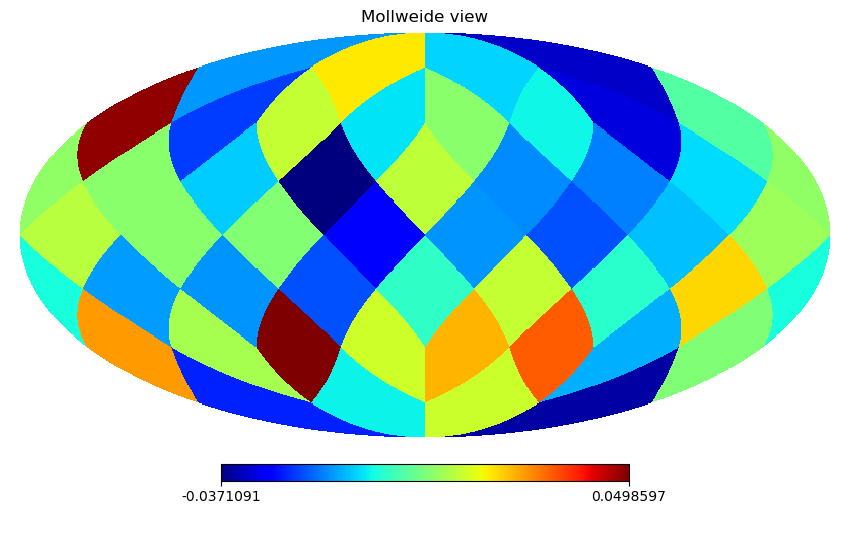}
         \caption{$n_s$}
    \end{subfigure}
    ~ 
    \begin{subfigure}[b]{0.45\textwidth}
        \includegraphics[trim={0 0 0 0.8cm},clip,width=\textwidth]{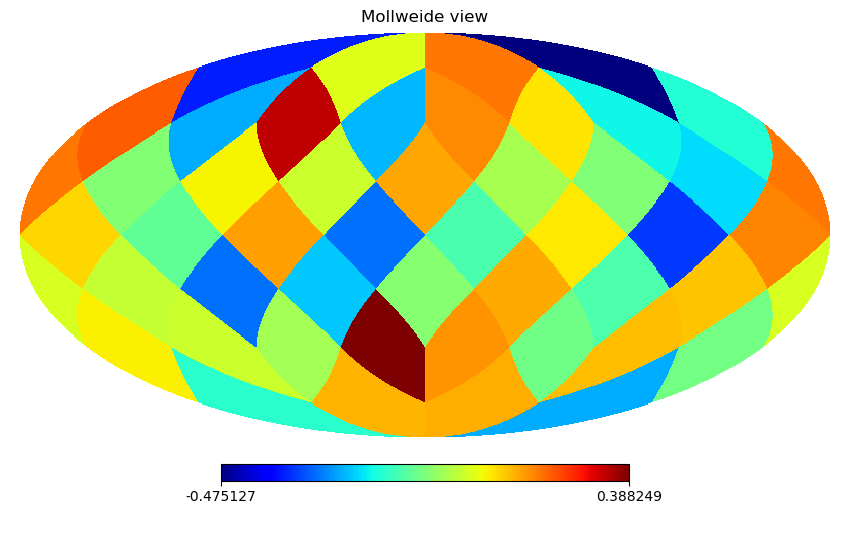}
         \caption{$A_L$}
    \end{subfigure}
       \caption{The azimuthally symmetric method mentioned in Eq.~\ref{eq:symmetricpatches} is implemented on simulated CMB sky to obtain six cosmological parameter maps for a patch radius of $16.5^\circ$.}\label{conv_sims}
 \end{figure}

\begin{figure}[H]
    \centering
    \begin{subfigure}[b]{0.45\textwidth}
        \includegraphics[trim={0 0 0 0.8cm},clip,width=\textwidth]{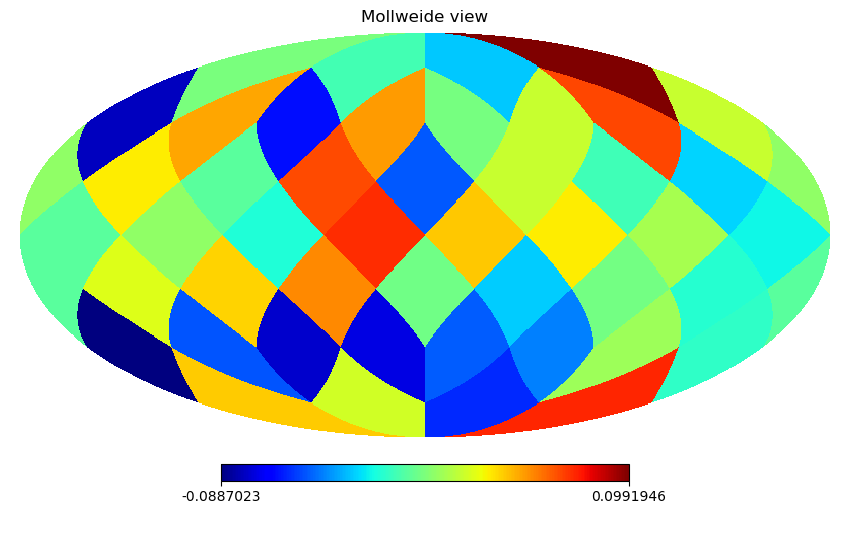}
        \caption{$A_s$}
    \end{subfigure}
    ~ 
    \begin{subfigure}[b]{0.45\textwidth}
    \includegraphics[trim={0 0 0 0.8cm},clip,width=\textwidth]{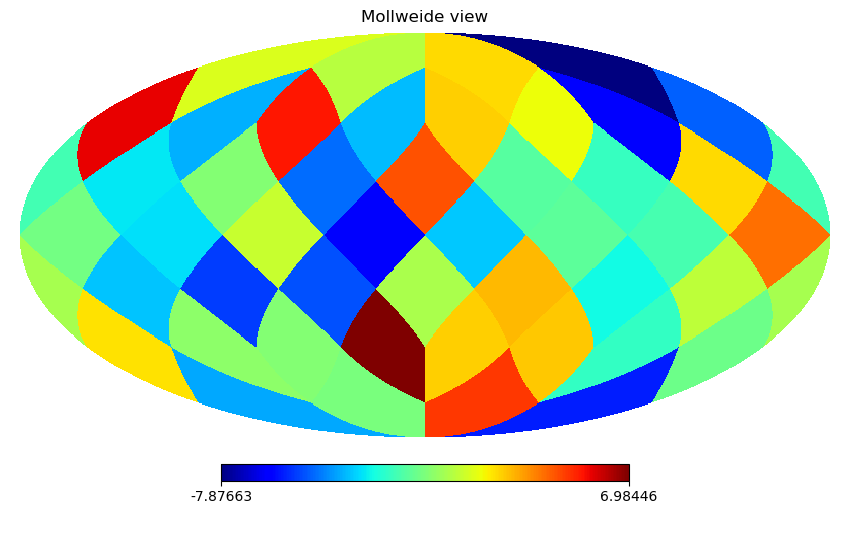}
        \caption{$H_0$}
    \end{subfigure}
    \begin{subfigure}[b]{0.45\textwidth}
   \includegraphics[trim={0 0 0 0.8cm},clip,width=\textwidth]{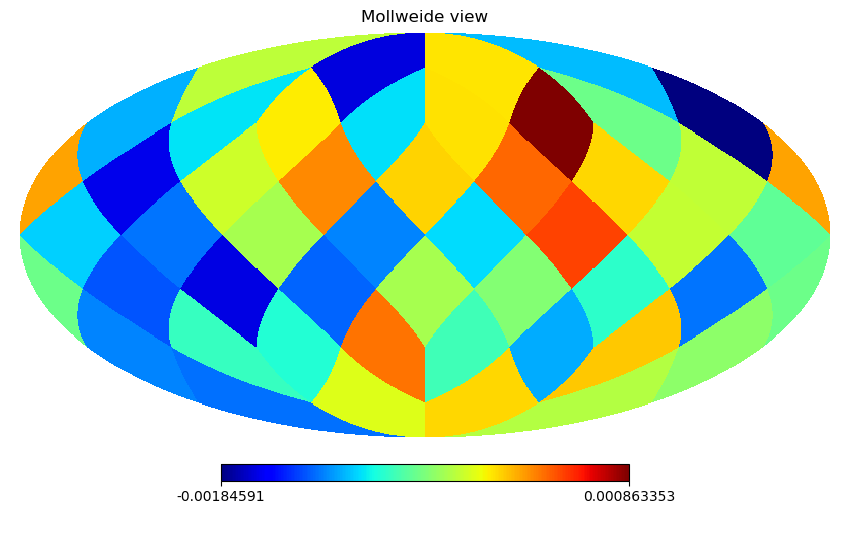}
        \caption{$O_bh^2$}
    \end{subfigure}
    ~ 
    \begin{subfigure}[b]{0.45\textwidth}
        \includegraphics[trim={0 0 0 0.8cm},clip,width=\textwidth]{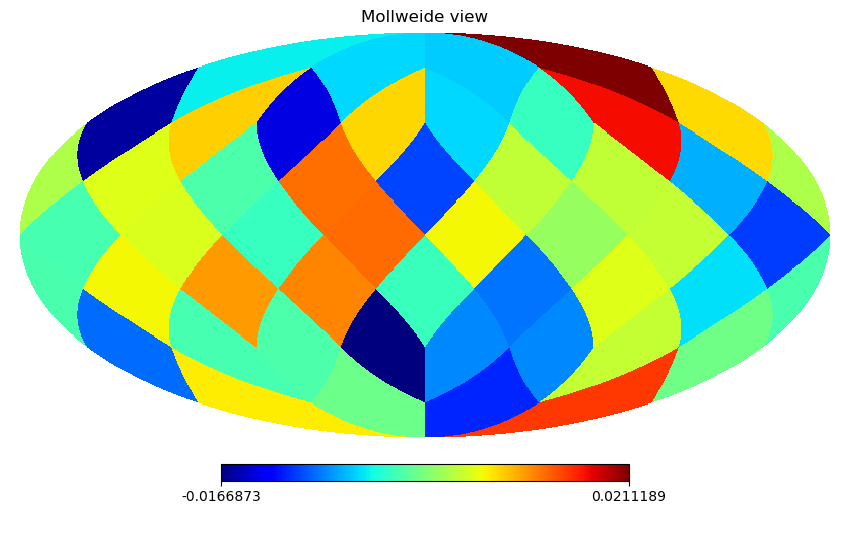}
         \caption{$O_ch^2$}
    \end{subfigure}
    \begin{subfigure}[b]{0.45\textwidth}
    \includegraphics[trim={0 0 0 0.8cm},clip,width=\textwidth]{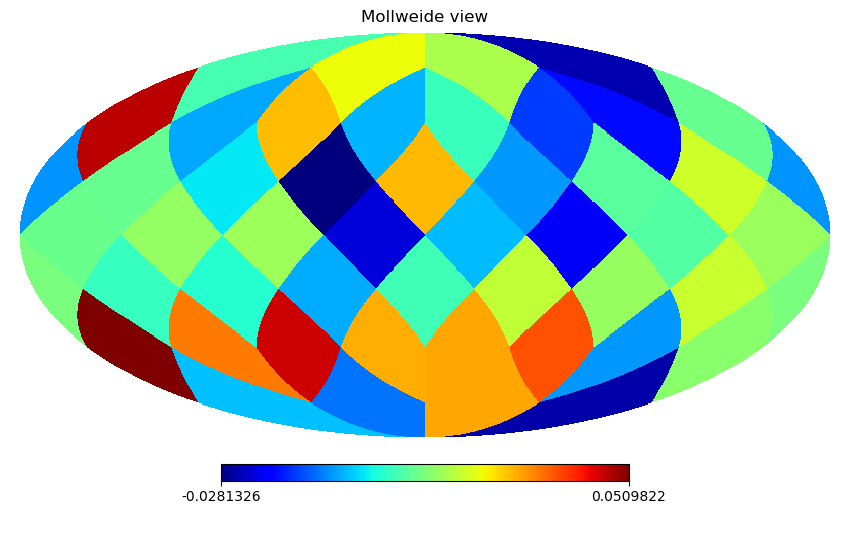}
         \caption{$n_s$}
    \end{subfigure}
    ~ 
    \begin{subfigure}[b]{0.45\textwidth}
        \includegraphics[trim={0 0 0 0.8cm},clip,width=\textwidth]{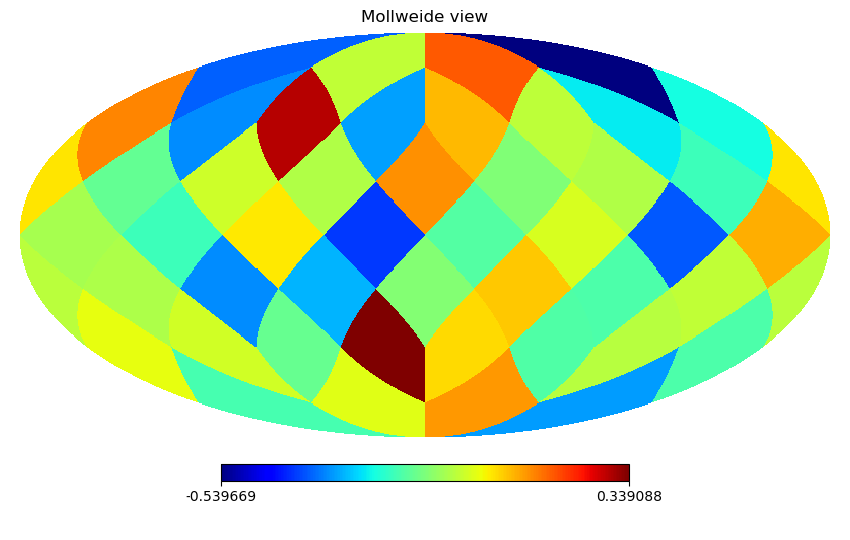}
         \caption{$A_L$}
    \end{subfigure}
       \caption{The arbitrary mask case, Eq.~\ref{eq:generalestimator} implemented on the same simulated CMB sky as in Fig.~\ref{conv_sims} to obtain six cosmological parameter maps for a patch radius of $16.5^\circ$. Note that while the approximation in Eq.~\eqref{eq:approx} produces minor changes, the binning of $\Delta l=20$ used in the arbitrary patch analysis means the maps are not expected to match exactly.}\label{patch_sims} \end{figure}

 The azimuthal symmetry method and the arbitrary mask method mentioned in Eq. \eqref{eq:symmetricpatches} and Eq.\eqref{eq:generalestimator}) respectively are now implemented for an ideal simulated map of a fixed realization. We  implemented these methods on the simulated CMB map obtained by HEALPix \cite{Gorski:2004by} using the best-fit Planck cosmological parameters \cite{Ade:2015xua}, to get a fluctuation map for six cosmological parameters ($A_s, n_s, \text{O}_bh^2, \text{O}_ch^2, \text{H}_0, A_L$).  
 We applied Eq. \eqref{eq:symmetricpatches} and Eq.\eqref{eq:generalestimator}) with a symmetric circular patch of radius $16.5^\circ$ and obtained the parameter maps as depicted in Fig.~\ref{conv_sims} and Fig.~\ref{patch_sims} respectively for both these cases. As expected from a simulated sky, all the fluctuations in the cosmological parameters are consistent with the best-fit value used in the analysis. We also obtain a cosmological parameter map in high resolution using azimuthally symmetric method. 
 A relative RMS difference in the values from both these methods are less than 1 sigma and are shown in Table~\ref{tab_1}. This indicates that both the method are accurate and giving us consistent results. The azimuthally symmetric method is significantly computationally faster than the other method. The details on the computational cost  are discussed in the Sec.~\ref{sec:computationalcost}.
   \begin{table*}
\centering
\caption{Root mean square differences between the arbitrary patch and the symmetric patch methods in units of the standard deviation. Note that  while the approximation in Eq.~\eqref{eq:approx} produces minor changes,  the binning of $\Delta l=20$ used in the arbitrary patch analysis means the results are not expected to agree exactly. }
\label{tab_1} 
\vspace{0.5cm}
\begin{tabular}{|p{2.5cm}|p{3.5cm}|}
\hline 
\centering Cosmological Parameters & \centering RMS fluctuations (in units of the standard deviation) \tabularnewline
\hline
\centering$A_s$ & \centering 0.33 \tabularnewline
\hline
\centering$H_0$ & \centering 0.26 \tabularnewline
\hline
\centering$O_bh^2$ & \centering 0.76 \tabularnewline
\hline
\centering$O_ch^2$ & \centering 0.26 \tabularnewline
\hline
\centering$n_s$ & \centering 0.53 \tabularnewline
\hline
\centering$A_L$ & \centering 0.28 \tabularnewline
\hline
\end{tabular}
\end{table*}
\subsection{Computational cost}\label{sec:computationalcost}
\,
\textbf{Arbitrary patches :} For the general case, Eq.~\ref{eq:generalestimator},  the computational cost is dominated by the spherical harmonic transforms to compute the pseudo-spectra and coupling matrices for all patches, ie 2 per patch. These computations can be done in parallel for all patches, enabling embarrassingly parallel computation on $n_{\text{patch}}$ nodes (in addition to any parallelization of the harmonic transforms).   The total number of operations is more than the symmetric case. For the example shown in Fig.~\ref{patch_sims} the method takes 83 minutes on a $3.5$ GHz CPU  for the six cosmological parameter maps.\\
\textbf{Azimuthally symmetric patches :} In this case, we need to perform  one transformation for the mask, three  spherical harmonic transformations for each parameter at the native resolution of the map  and on final transformation to the patch resolution. So, for a set of $n_{\theta}$ parameters, we need to perform $3n_{\theta}+1$ high resolution transformations. As a result, for a set of six cosmological parameters, we need to perform 19 high resolution transformations. For a parallel computational process, we can distribute the job to $n_{\theta}$ computers and each computer needs to perform 3 high resolution transformations and one low resolution one. For larger resolution of the parameter map, and hence smaller and more numerous patches, this method rapidly becomes much faster general patch shape. Even for the example shown in Fig.~\ref{conv_sims} with only 48 parameter patches the method takes  10 minutes on a 3.5 GHz CPU for all six full-sky cosmological parameter maps, a speed-up of a factor of 8.

\section{Implementation on Planck SMICA map of CMB}\label{results}

\subsection{Procedure}\label{procedure}
The implementation of the method formulated in Sec.~\ref{formalism} requires calculating three essential ingredients: the  covariance matrix for the angular power spectra ($\bm{K_{\delta C}}$), the derivative matrix $\bm{D}$ and the estimate of the difference in the angular power spectra of each patches from the fiducial value of the angular power spectra $\bm{c}$. 

\begin{figure}[h]
\centering
\includegraphics[trim={0 0 0 1.7cm},clip, width=4.5in,keepaspectratio=true]{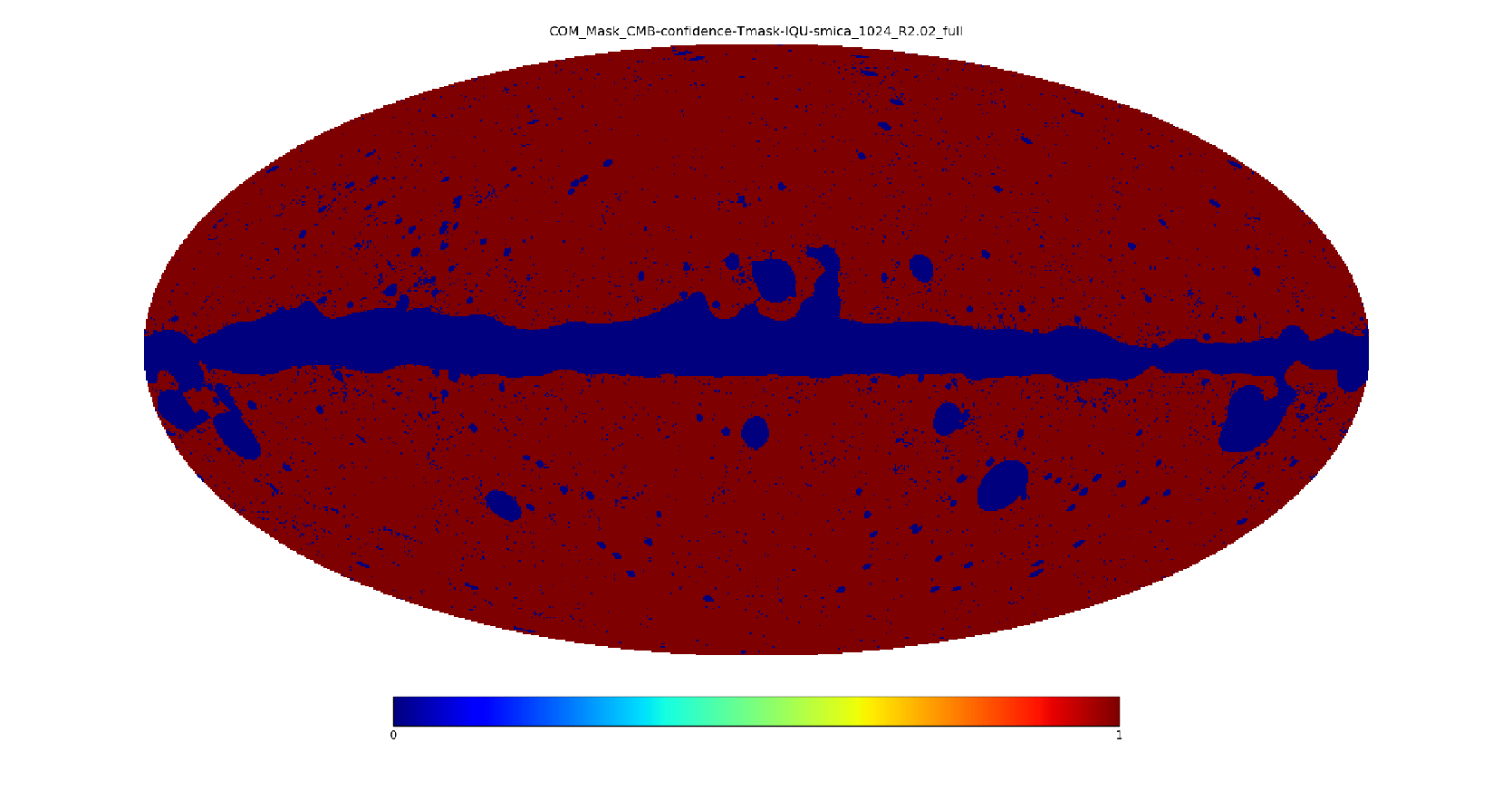}
\caption{SMICA mask used in the analysis to obtain the variation in the cosmological parameters.}\label{smica-mask}
\end{figure}

\textbf{Calculation of $\bm{c}$}:  On the nside $=2048$ resolution maps of Planck, we applied the Planck SMICA confidence mask as depicted in Fig.~\ref{smica-mask}. We divide the sky into a low resolution NSIDE$=2$ map having $48$ patches. We then obtain the angular power spectrum $C_l$  from each of these patches using HEALPix \cite{Gorski:2004by}. The $C_l$ from the masked sky for each of the patches are corrected using MASTER \cite{Hivon:2001jp} to take care of the effects of the partial sky. Then the difference between the estimated $C_l (\hat n)$ from each patch centered in the direction ($\hat n$) and the fiducial $C_l$ are obtained for each patch. The fiducial $C_l$ is obtained using  CAMB \cite{Lewis:1999bs} by using the fiducial cosmological parameters from Planck \cite{Ade:2015xua}.

\textbf{Calculation of $\bm{K_{\delta C}}$}: The binned spectra   are taken to have covariance   \cite{Hivon:2001jp}
\begin{equation}
K_{\delta C}= \frac{2}{(2l+1)\Delta l f_{sky}} (C_l + N_l)^2, 
\end{equation}
where $N_l$ is the instrumental noise in the SMICA map. The $f_{sky}$ for each patch is different due to different contributions from the galactic mask. The error bar for each patch is therefore different, most notably along the galactic plane where patches have a smaller sky fraction and hence bigger error bars in comparison to the patches at  higher latitudes.

\textbf{Calculation of $\bm{D}$}: The $\bm{D}$ matrix captures the derivative of the angular power spectra evaluated at the fiducial value of the parameters $C_l$. We evaluate the angular power spectra $C_l$ from CAMB \cite{Lewis:1999bs} by varying each parameter individually and keeping all other parameters fixed at the fiducial cosmological parameters from Planck-2015 \cite{Ade:2015xua}. We then obtain the numerical derivative of the angular power spectra at every values of $l$, and construct a $n_{\theta}\times l$ dimension matrix, where $n_{\theta}$ indicates the  number of parameters.  



\subsection{Parameter variation in Planck SMICA map}
\begin{figure}[H]
    \centering
    \begin{subfigure}[b]{0.45\textwidth}
\includegraphics[trim={0 0 0 0.8cm},clip,width=\textwidth]{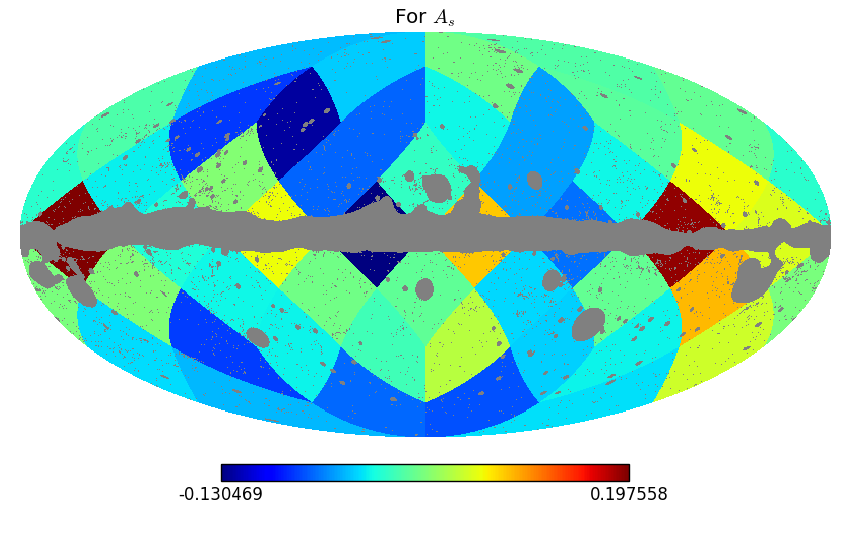}
        \caption{$A_s$}
    \end{subfigure}
    ~ 
    \begin{subfigure}[b]{0.45\textwidth}
        \includegraphics[trim={0 0 0 0.8cm},clip,width=\textwidth]{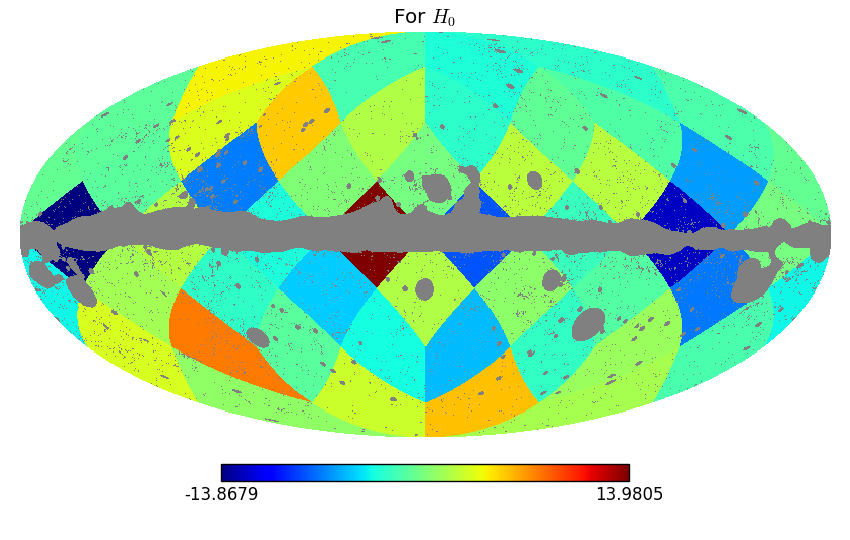}
        \caption{$H_0$}
    \end{subfigure}
    \begin{subfigure}[b]{0.45\textwidth}
    \includegraphics[trim={0 0 0 0.8cm},clip,width=\textwidth]{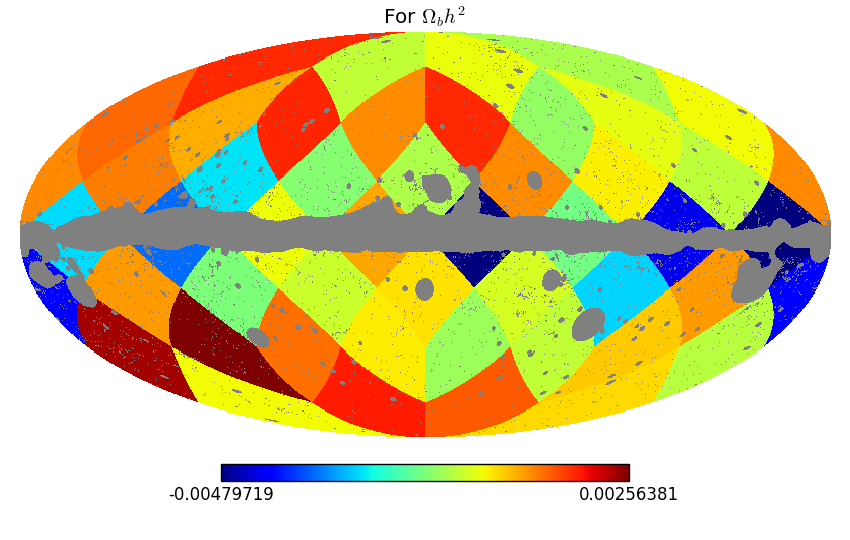}
        \caption{$O_bh^2$}
    \end{subfigure}
    ~ 
    \begin{subfigure}[b]{0.45\textwidth}
        \includegraphics[trim={0 0 0 0.8cm},clip,width=\textwidth]{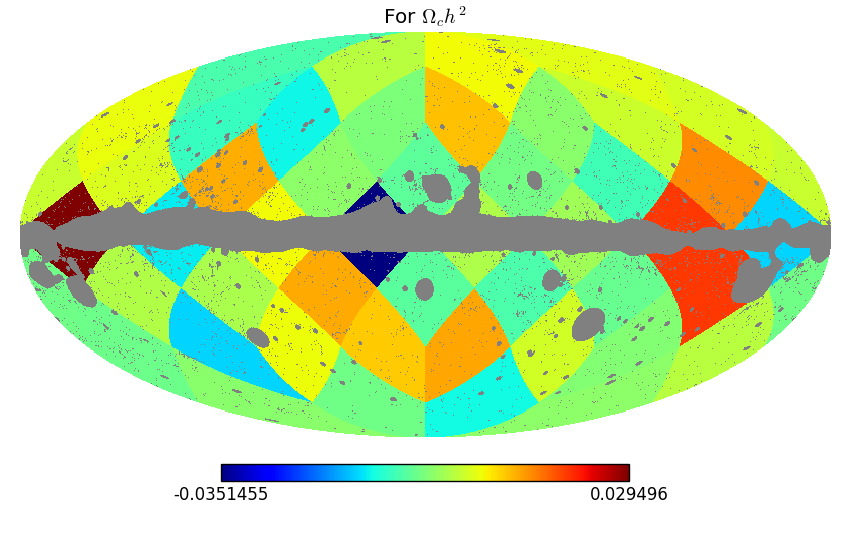}
        \caption{$O_ch^2$}
    \end{subfigure}
    \begin{subfigure}[b]{0.45\textwidth}
 \includegraphics[trim={0 0 0 0.8cm},clip,width=\textwidth]{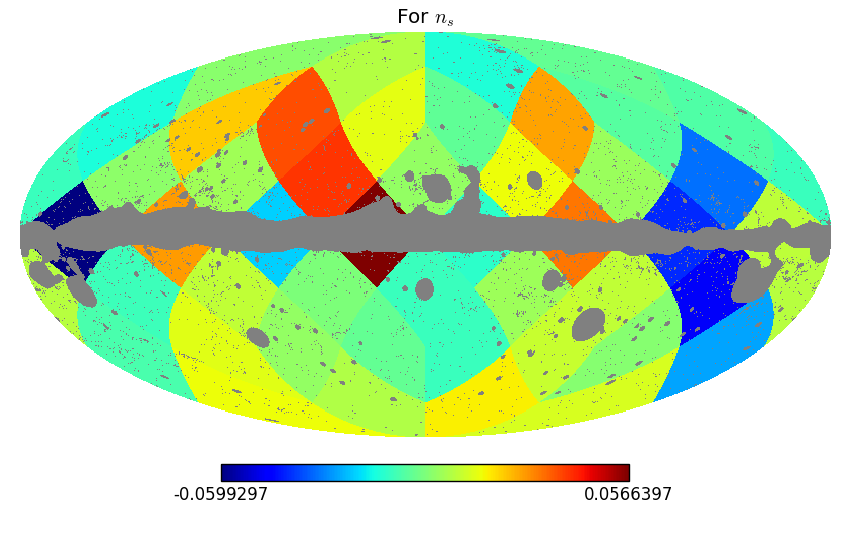}
        \caption{$n_s$}
    \end{subfigure}
    ~ 
    \begin{subfigure}[b]{0.45\textwidth}
\includegraphics[trim={0 0 0 0.8cm},clip,width=\textwidth]{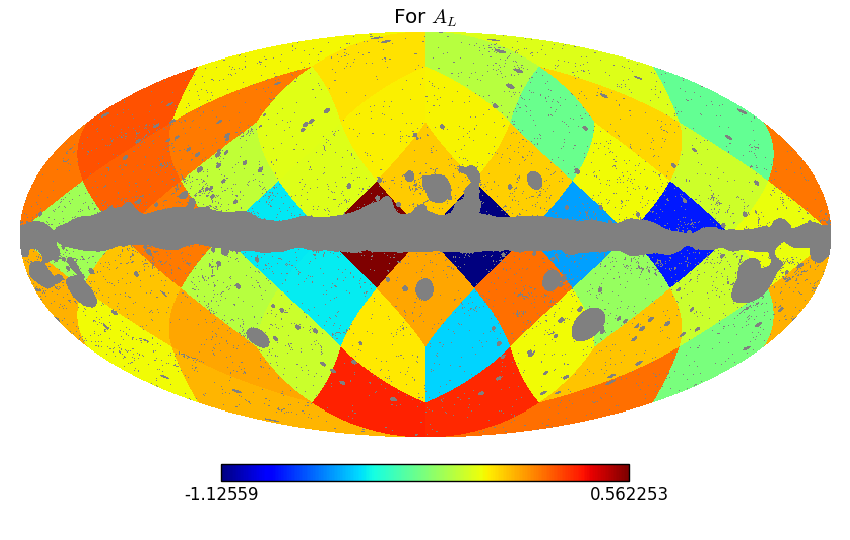}
        \caption{$A_L$}
    \end{subfigure}

   \caption{The spatial variation of six cosmological parameters from the Planck SMICA HM1$\times$ HM2 map are depicted along with the galactic mask used in the analysis. These results are obtained for the bin size of $\Delta l=50$ using CMB multipoles $[50, 1300]$.}\label{para-1}
 \end{figure}

\begin{figure}[H]
    \centering
    \begin{subfigure}[b]{0.45\textwidth}
        \includegraphics[trim={0 0 0 0.8cm},clip,width=\textwidth]{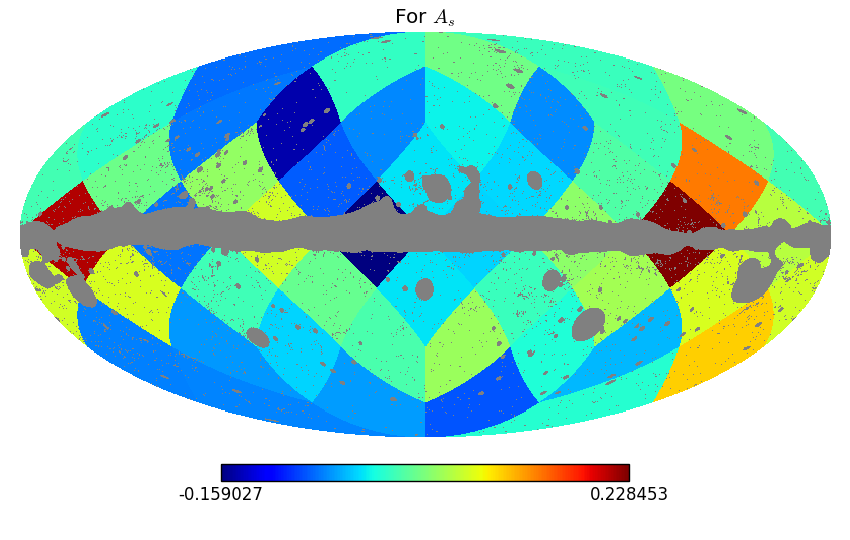}
        \caption{$A_s$}
    \end{subfigure}
    ~ 
    \begin{subfigure}[b]{0.45\textwidth}
    \includegraphics[trim={0 0 0 0.8cm},clip,width=\textwidth]{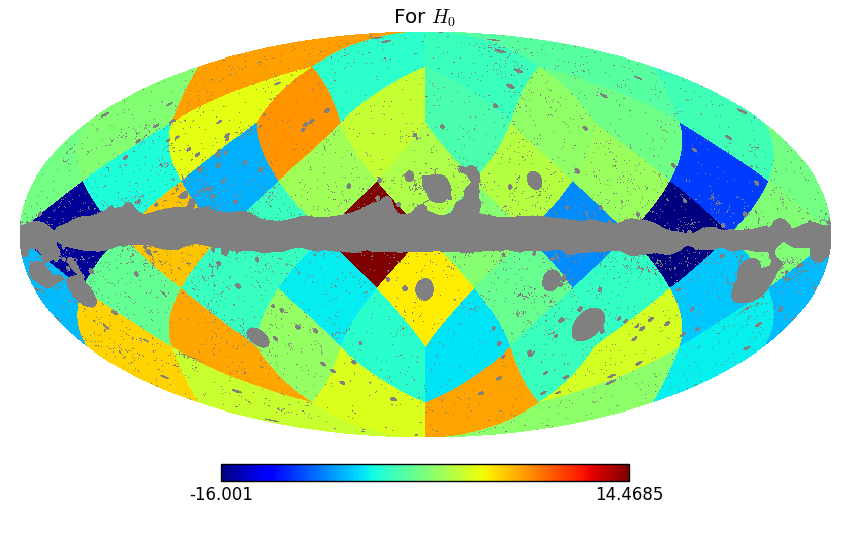}
        \caption{$H_0$}
    \end{subfigure}
    \begin{subfigure}[b]{0.45\textwidth}
    \includegraphics[trim={0 0 0 0.8cm},clip,width=\textwidth]{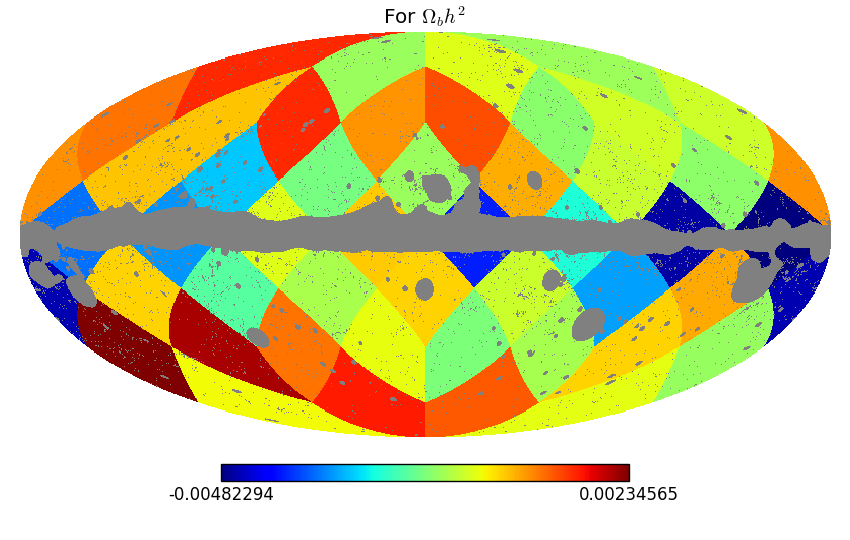}
        \caption{$O_bh^2$}
    \end{subfigure}
    ~ 
    \begin{subfigure}[b]{0.45\textwidth}
        \includegraphics[trim={0 0 0 0.8cm},clip,width=\textwidth]{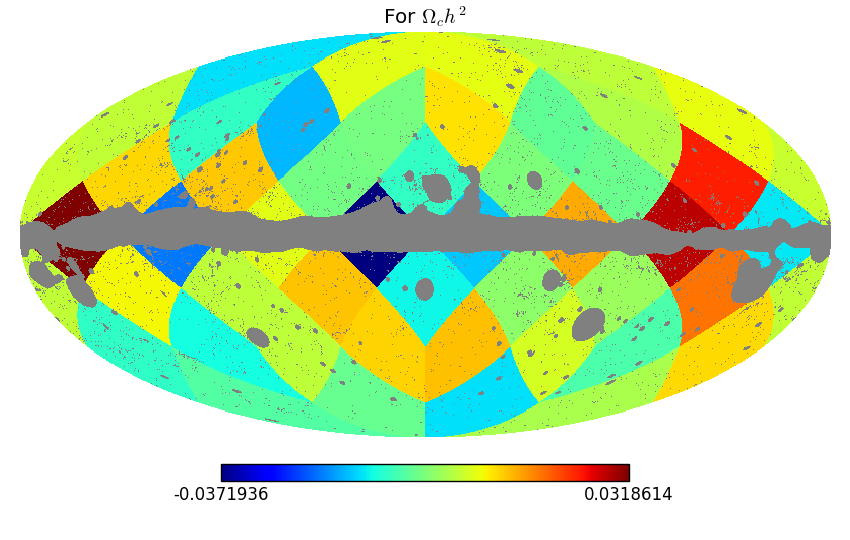}
        \caption{$O_ch^2$}
    \end{subfigure}
    \begin{subfigure}[b]{0.45\textwidth}
    \includegraphics[trim={0 0 0 0.8cm},clip,width=\textwidth]{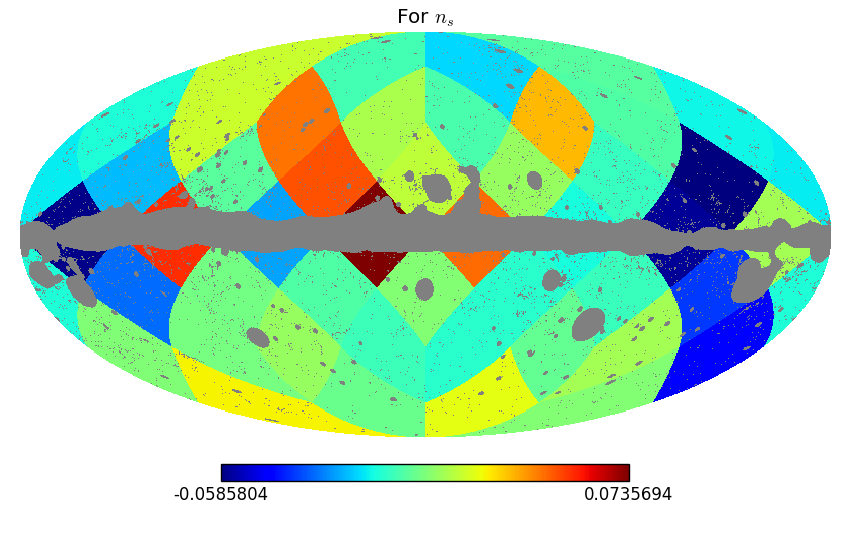}
        \caption{$n_s$}
    \end{subfigure}
    ~ 
    \begin{subfigure}[b]{0.45\textwidth}
    \includegraphics[trim={0 0 0 0.8cm},clip,width=\textwidth]
    {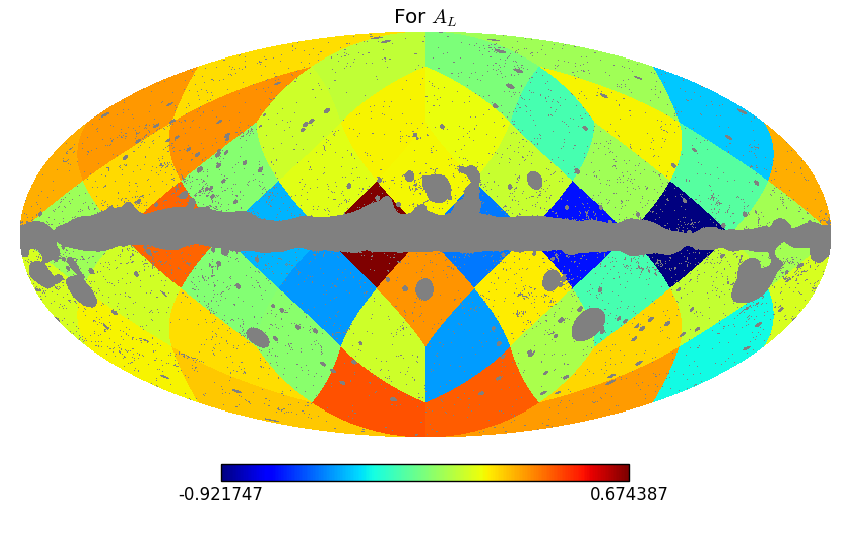}
        \caption{$A_L$}
    \end{subfigure}

   \caption{The spatial variation of six cosmological parameters from the Planck SMICA HM1$\times$ HM2 map are depicted along with the galactic mask used in the analysis. These results are obtained for the bin size of $\Delta l=20$ using CMB multipoles $[20, 1300]$.}\label{para-2}
 \end{figure}

\begin{figure}[H]
    \centering
    \begin{subfigure}[b]{0.45\textwidth}
        \includegraphics[width=\textwidth]{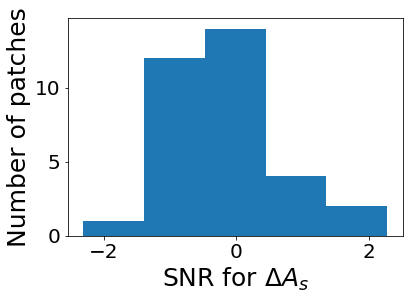}
    \end{subfigure}
    ~ 
    \begin{subfigure}[b]{0.45\textwidth}
        \includegraphics[width=\textwidth]{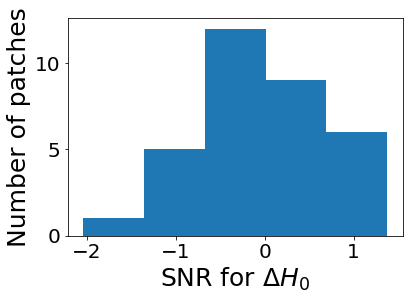}
    \end{subfigure}
    \begin{subfigure}[b]{0.45\textwidth}
    \includegraphics[width=\textwidth]{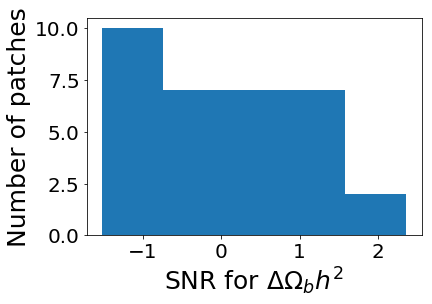}
    \end{subfigure}
    ~ 
    \begin{subfigure}[b]{0.45\textwidth}
        \includegraphics[width=\textwidth]{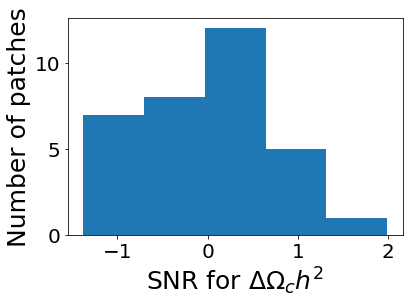}
    \end{subfigure}
    ~ 
    \begin{subfigure}[b]{0.45\textwidth}
        \includegraphics[width=\textwidth]{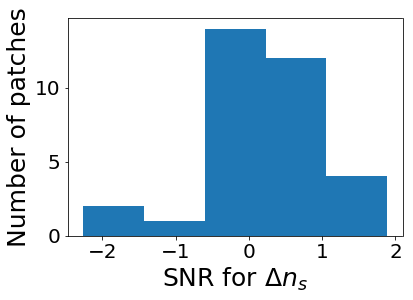}
        \end{subfigure}
    \begin{subfigure}[b]{0.45\textwidth}
 \includegraphics[width=\textwidth]{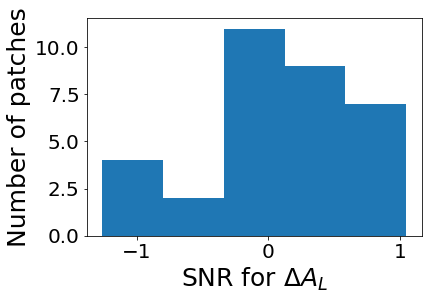}
    \end{subfigure}

   \caption{The distribution of SNR for cosmological parameters for different patches from Planck SMICA HM1$\times$ HM2. This is the case with bin width $\Delta l=20$ used in the Master algorithm and CMB multipole range [20, 1300]. The distribution does not include the patches near the galactic plane having $f_{sky}\leq0.017$.}\label{hist20}
 \end{figure}

Using all the tools mentioned in the previous section, we obtain the variation of cosmological parameters in the sky for six cosmological parameters  ($A_s,$ $n_s,$ $\text{O}_bh^2,$ $\text{O}_ch^2,$ $\text{H}_0,$ and $A_L$) for two different choices of $\Delta l=20$ \& $50$. The value of  all other parameters are kept fixed at the fiducial Planck best-fit values \cite{Ade:2015xua}  with $\tau =0.058$ \cite{Aghanim:2016fhp}. 
The results of parameter variation are depicted in Fig.~\ref{para-1} and Fig.~\ref{para-2}  for $\Delta l=50$ and $20$ respectively. These results show the deviation from the Planck fiducial values. We also overplot the galactic mask on the recovered parameters map to show the contribution of the galactic mask in each patch. The largest deviation is observed for certain patches in the galactic plane, which is anyway not robust for cosmological parameters and not used in the Planck likelihood. The parameters at high latitudes show negligible variation from the fiducial parameter values.
To indicate the deviations more clearly, we make histogram plots in Fig.~\ref{hist20} for the deviation of the parameters, inverse weighted with the square-root of the corresponding diagonal element of the inverse of the Fisher matrix for each patch. 
The histogram plots for the cosmological parameters are obtained for all the patches except the patches in the galactic plane that have sky fraction of less than 1.7 \%. The plots indicate that there is no significant deviation in any of the patches and  fluctuations from the fiducial parameters in most of the patches are less than $1\sigma$, ie underdispersed compared to expectations. This underdispersion arises because the estimators are correlated between patches. Even though they are calculated from different patches, they share the same realisation of the underlying CMB signal.

\begin{figure}[h]
\centering
\begin{subfigure}{0.45\textwidth}
\includegraphics[width=\textwidth]{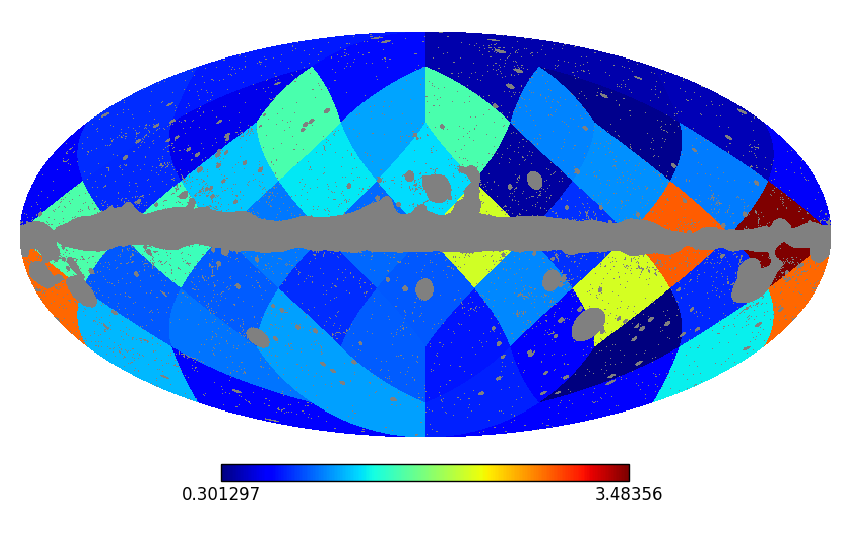}
\captionsetup{singlelinecheck=on,justification=raggedright}
\caption{For the multipole range [20, 1300]}\label{fig-chisqhigh}
\end{subfigure}
\begin{subfigure}{0.45\textwidth}
\includegraphics[width=\textwidth]{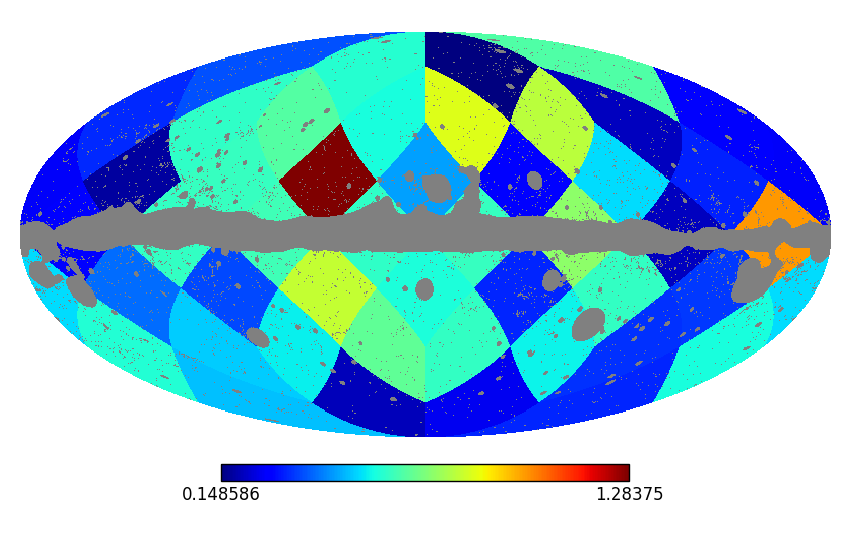}
\captionsetup{singlelinecheck=on,justification=raggedright}
\caption{For the multipole range [20, 520]}\label{fig-chisqlow}
\end{subfigure}
\caption{$\chi$ map for parameters (as expressed in Eq. \eqref{chisq}) for HM1 $\times$ HM2 SMICA CMB sky with bin width of $\Delta l=20$ used in the Master algorithm.}\label{fig-chisq}
\end{figure}

To quantify the total patch-to-patch variation in cosmological parameters, we estimate the reduced chi-square defined as
\begin{align}\label{chisq}
\begin{split}
\chi (\hat n)= \sqrt{\frac{\sum_{ij}\theta_i(\hat n)F_{ij}\theta_j(\hat n)}{\text{Number of parameters}}}.
\end{split}
\end{align}
This quantity captures the total deviation in the parameter values and is depicted in Fig.~\ref{fig-chisq}. It clearly indicates that the Planck SMICA map at high galactic latitude  is consistent with the global value of the parameters derived from the Planck likelihood \cite{Ade:2015xua}. This map captures the complete SNR of the deviation observed over all the parameters considered here. This extreme variation in a few patches is evident only for the case shown in Fig.~\ref{fig-chisqhigh}, which considers a wide multipole range (20-1300) and reduces significantly for the estimation which considers a lower multipole range (20-520) as shown in Fig.~\ref{fig-chisqlow}. 

\subsection{Comparison of Planck SMICA results with CMB simulations}\label{comp-sims}
To scrutinize this in detail and evaluate the statistical significance of the parameter variations we saw in the SMICA map, we applied our method to 100 simulated temperature maps.
The $\chi (\hat n)$ values from $100$ simulations (denoted by $\chi_s (\hat n)$) for each patch are listed and are arranged in an ascending order, which can be named as $\mathcal{L} (\hat n)$. Then the elements in the list $\mathcal{L} (\hat n)$ are grouped in sets of ten and each set is indexed with a \textit{group index} whose value varies from one to ten.  For any particular patch centered in the direction $\hat n_p$, having the smallest $\chi_s (\hat n_p)$ value is the first element of the list $\mathcal{L} ^{p}$ and hence carries the $\textit{group index} =1$. Whereas the largest $\chi_s (\hat n_p)$ value is the last element of the list $\mathcal{L} ^{p}$ and carries the $\textit{group index} =10$. 

The value of $\chi (\hat n)$ from Planck SMICA map (denoted by $\chi_d(\hat n)$) can be compared with the ordered list $\mathcal{L} (\hat n)$ for every sky  patches and we can obtain the \textit{group index} which is associated with the value of $\chi_d$.  
In this method of classification,  the value of \textit{group index} signifies the rank of the $\chi_d (\hat n)$ value with respect to $\chi_s (\hat n)$. For example, if $\chi_d$ value for a particular patch is having a $\textit{group index}=j$, then there are at-least $10\times(10-j)$ simulations having $\chi_s$ value which are more than $\chi_d$ for that patch. If the value of $\chi_d$ is smaller (or greater) than the lowest (or highest) value of the list $\mathcal{L}$, then we show it by the down (or up) arrow. 

\begin{figure}[h]
\centering
\begin{subfigure}{0.7\textwidth}
\includegraphics[width=\textwidth]{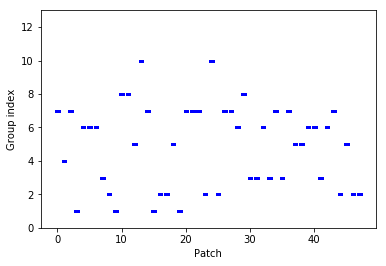}
\caption{For the multipole range [20, 520]}\label{gp:520}
\end{subfigure}
\begin{subfigure}{0.7\textwidth}
\includegraphics[width=\textwidth]{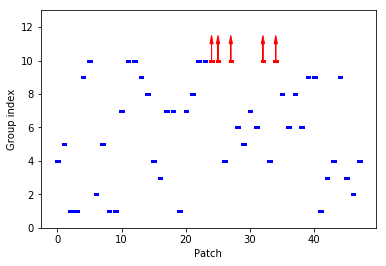}
\caption{For the multipole range [20, 1300]}\label{gp:1300}
\end{subfigure}
\captionsetup{singlelinecheck=on,justification=raggedright}
\caption{The \textit{group index} of $\chi_p$ values from the Planck SMICA map are indicated for each sky patches for bin width $\Delta l=20$ for (a) $l_{max}= 520$ and (b) $l_{max}= 1300$.  The red arrows in the plot indicate the patches for which the values of $\chi_p$ is larger than the values observed in the simulations. All the patches except for five that intersect the galactic plane (colored yellow to red in Fig.~\ref{fig-chisq}) are consistent with the best-fit Planck parameters. 
}\label{fig:group-index}
\end{figure}

Using this prescription, we obtain the \textit{group index} for all the sky patches and depict them in Fig.~\ref{gp:520} and Fig.~\ref{gp:1300} for $l_{max} = 520$ and $l_{max}=1300$ respectively.
Fig.~\ref{gp:520} shows that for all the patches with $l_{max}=520$, Planck SMICA map is consistent with the values obtained from the simulations and none of the deviations in the cosmological parameters are significant. However, for $l_{max}=1300$ depicted in Fig.~\ref{gp:1300}, a few patches near the galactic plane show higher deviations and do not match with the values of $\chi_s$ obtained from $100$ simulations. These five patches are indicated by red arrows in Fig.~\ref{gp:1300} and can also be identified from the $\chi$ map in Fig.~\ref{fig-chisq}. All other patches are consistent with the statistically isotropic simulations and do not show any significant departures. 

\section{Conclusions}\label{conc}
We described a fast algorithm to estimate the local variation of cosmological parameters from their fiducial value. In this method, we implement a local, optimal, quadratic filter that projects the data onto fields whose variances contain the cosmological parameter information. The full algorithm is described in  Sec.~\ref{formalism}. The advantage of this method is that it does not require  costly MCMC analysis to obtain the deviations in the cosmological parameters. This makes the algorithm very useful to apply to any large data set. We implement it on the Planck SMICA-2015 temperature map to obtain the direction dependence of the six standard cosmological parameters ($A_s, n_s, \text{O}_bh^2, \text{O}_ch^2, \text{H}_0, A_L$). The parameters maps are  shown in Fig.~\ref{para-1} and Fig.~\ref{para-2} for two different choices of power spectrum binning $\Delta l=50$ and $\Delta l=20$ respectively.  The maximum variation is evident in  $\Omega_bh^2$ at approximately $2\sigma$ in two patches. 
 In particular, parameters like $H_0$, which shows some discrepancy compared to other data sets, and $A_L$ which shows some tension with the expected value from standard model,  do not show significant directional dependence. Our estimate shows that both these parameters are within the   $2\sigma$ variation for all the patches. We find that $A_L$ shows the least variations between patches in comparison to all other parameters.  We compare our results with simulated uncontaminated CMB maps to get a better understanding of our SMICA results. We mask the simulated maps with the SMICA mask as shown in Fig.~\ref{smica-mask}. Our simulation results  indicate that the cosmological parameters based on the Planck SMICA-2015 temperature map do not exhibit higher variation than what  is to be expected based on statistical fluctuations except in some patches on the galactic plane that are not used in the reference Planck analysis.  
On a $\chi$ map for SMICA (Fig.~\ref{fig-chisq}) for the range of multipoles [20, 1300], we see that the parameter values in a few patches in the galactic plane exhibit a higher departure than our 100 simulations. But the patches at high galactic latitudes do not show any strong deviation in comparison to statistically isotropic simulations. 

While the main motivation of our work was to provide a tool for checking for parameter variations that might visually correlate with known sources of systematics (from the Galaxy or more locally) our method can of course be used to explore a possible breaking of statistical isotropy. If a suggestive pattern were seen, it might motivate an in-depth analysis with a inference that is optimised for a particular physics model of isotropy breaking. We find no obvious indications of systematic  parameter variations  in the Planck SMICA map. A cross-correlation between the cosmological parameter map obtained from Planck SMICA-2015 and the results from the supernovae data by previous studies \cite{Javanmardi:2015sfa, Carvalho:2015lqd, Carvalho:2016aah} can be useful to scrutinize any common direction dependence in the two different data sets.

In summary, we present an efficient algorithm to check for any directional dependence of the cosmological parameters. Due to the computational efficiency of our method, it can be used for large data sets, to check for the effects of contaminations from unknown systematics on the estimated cosmological parameters.  This method is applicable not only to CMB data set but adapts to other data sets, such as  galaxy surveys, weak-lensing shear measurements, and $21$ cm data   with minimal effort. 

Future observational campaigns promise to deliver robust measurements of cosmological parameters despite  challenging foreground and noise contaminations. The methods we present in this paper add an efficient diagnostic tool to assess how these final science products are affected by potential instrumental or astrophysical systematics.\\

\textbf{Acknowledgements}
This work has been done within the Labex ILP (reference ANR-10-LABX-63) part of the Idex SUPER, and received financial state aid managed by the Agence Nationale de la Recherche, as part of the programme Investissements d'avenir under the reference ANR-11-IDEX-0004-02. This work is supported by the Simons Foundation. The author acknowledges the use of CAMB, HEALPix and MASTER algorithm in this analysis.

\bibliographystyle{unsrtads}
\bibliography{Direction-dependent-parameter-estimation-paper-final}

\begin{thebibliography}{10}

\bibitem{1965ApJ...142..419P}
A.~A. {Penzias} and R.~W. {Wilson}.
\newblock {A Measurement of Excess Antenna Temperature at 4080 Mc/s.}
\newblock {\em \apj}, 142:419--421, July 1965.
\newblock \href {http://dx.doi.org/10.1086/148307} {\path{[DOI]}},
  {\small[\href{http://adsabs.harvard.edu/abs/1965ApJ...142..419P}{ADS}]}.

\bibitem{firas}
D.~J. {Fixsen}, E.~S. {Cheng}, J.~M. {Gales}, J.~C. {Mather}, R.~A. {Shafer},
  and E.~L. {Wright}.
\newblock {The Cosmic Microwave Background Spectrum from the Full COBE FIRAS
  Data Set}.
\newblock {\em \apj}, 473:576, 1996.
\newblock \href {http://dx.doi.org/10.1086/178173} {\path{[DOI]}},
  {\small[\href{http://adsabs.harvard.edu/abs/1996ApJ...473..576F}{ADS}]}.

\bibitem{firasan}
D.~J. {Fixsen}, G.~{Hinshaw}, C.~L. {Bennett}, and J.~C. {Mather}.
\newblock {The Spectrum of the Cosmic Microwave Background Anisotropy from the
  Combined COBE FIRAS and DMR Observations}.
\newblock {\em \apj}, 486:623--628, September 1997.
\newblock \href {http://arxiv.org/abs/astro-ph/9704176}
  {\path{arXiv:astro-ph/9704176}}, \href {http://dx.doi.org/10.1086/304560}
  {\path{[DOI]}},
  {\small[\href{http://adsabs.harvard.edu/abs/1997ApJ...486..623F}{ADS}]}.

\bibitem{fm2002}
D.~J. {Fixsen} and J.~C. {Mather}.
\newblock {The Spectral Results of the Far-Infrared Absolute Spectrophotometer
  Instrument on COBE}.
\newblock {\em \apj}, 581:817--822, 2002.
\newblock \href {http://dx.doi.org/10.1086/344402} {\path{[DOI]}},
  {\small[\href{http://adsabs.harvard.edu/abs/2002ApJ...581..817F}{ADS}]}.

\bibitem{cobe}
E.~L. {Wright}, C.~L. {Bennett}, K.~{Gorski}, G.~{Hinshaw}, and G.~F. {Smoot}.
\newblock {Angular Power Spectrum of the Cosmic Microwave Background Anisotropy
  seen by the COBE DMR}.
\newblock {\em \apjl}, 464:L21, 1996.
\newblock \href {http://dx.doi.org/10.1086/310073} {\path{[DOI]}},
  {\small[\href{http://adsabs.harvard.edu/abs/1996ApJ...464L..21W}{ADS}]}.

\bibitem{Bennett:2012zja}
C.~L. {Bennett}, D.~{Larson}, J.~L. {Weiland}, N.~{Jarosik}, G.~{Hinshaw},
  N.~{Odegard}, K.~M. {Smith}, R.~S. {Hill}, B.~{Gold}, M.~{Halpern},
  E.~{Komatsu}, M.~R. {Nolta}, L.~{Page}, D.~N. {Spergel}, E.~{Wollack},
  J.~{Dunkley}, A.~{Kogut}, M.~{Limon}, S.~S. {Meyer}, G.~S. {Tucker}, and
  E.~L. {Wright}.
\newblock {Nine-year Wilkinson Microwave Anisotropy Probe (WMAP) Observations:
  Final Maps and Results}.
\newblock {\em \apjs}, 208:20, October 2013.
\newblock \href {http://arxiv.org/abs/1212.5225} {\path{arXiv:1212.5225}},
  \href {http://dx.doi.org/10.1088/0067-0049/208/2/20} {\path{[DOI]}},
  {\small[\href{http://adsabs.harvard.edu/abs/2013ApJS..208...20B}{ADS}]}.

\bibitem{Adam:2015rua}
{Planck Collaboration}, R.~{Adam}, P.~A.~R. {Ade}, N.~{Aghanim}, Y.~{Akrami},
  M.~I.~R. {Alves}, F.~{Arg{\"u}eso}, M.~{Arnaud}, F.~{Arroja}, M.~{Ashdown},
  and et~al.
\newblock {Planck 2015 results. I. Overview of products and scientific
  results}.
\newblock {\em \aap}, 594:A1, September 2016.
\newblock \href {http://arxiv.org/abs/1502.01582} {\path{arXiv:1502.01582}},
  \href {http://dx.doi.org/10.1051/0004-6361/201527101} {\path{[DOI]}},
  {\small[\href{http://adsabs.harvard.edu/abs/2016A%26A...594A...1P}{ADS}]}.

\bibitem{Hanson:2013hsb}
D.~{Hanson}, S.~{Hoover}, A.~{Crites}, P.~A.~R. {Ade}, K.~A. {Aird}, J.~E.
  {Austermann}, J.~A. {Beall}, A.~N. {Bender}, B.~A. {Benson}, L.~E. {Bleem},
  J.~J. {Bock}, J.~E. {Carlstrom}, C.~L. {Chang}, H.~C. {Chiang}, H.-M. {Cho},
  A.~{Conley}, T.~M. {Crawford}, T.~{de Haan}, M.~A. {Dobbs}, W.~{Everett},
  J.~{Gallicchio}, J.~{Gao}, E.~M. {George}, N.~W. {Halverson},
  N.~{Harrington}, J.~W. {Henning}, G.~C. {Hilton}, G.~P. {Holder}, W.~L.
  {Holzapfel}, J.~D. {Hrubes}, N.~{Huang}, J.~{Hubmayr}, K.~D. {Irwin},
  R.~{Keisler}, L.~{Knox}, A.~T. {Lee}, E.~{Leitch}, D.~{Li}, C.~{Liang},
  D.~{Luong-Van}, G.~{Marsden}, J.~J. {McMahon}, J.~{Mehl}, S.~S. {Meyer},
  L.~{Mocanu}, T.~E. {Montroy}, T.~{Natoli}, J.~P. {Nibarger}, V.~{Novosad},
  S.~{Padin}, C.~{Pryke}, C.~L. {Reichardt}, J.~E. {Ruhl}, B.~R. {Saliwanchik},
  J.~T. {Sayre}, K.~K. {Schaffer}, B.~{Schulz}, G.~{Smecher}, A.~A. {Stark},
  K.~T. {Story}, C.~{Tucker}, K.~{Vanderlinde}, J.~D. {Vieira}, M.~P. {Viero},
  G.~{Wang}, V.~{Yefremenko}, O.~{Zahn}, and M.~{Zemcov}.
\newblock {Detection of B-Mode Polarization in the Cosmic Microwave Background
  with Data from the South Pole Telescope}.
\newblock {\em Physical Review Letters}, 111(14):141301, October 2013.
\newblock \href {http://arxiv.org/abs/1307.5830} {\path{arXiv:1307.5830}},
  \href {http://dx.doi.org/10.1103/PhysRevLett.111.141301} {\path{[DOI]}},
  {\small[\href{http://adsabs.harvard.edu/abs/2013PhRvL.111n1301H}{ADS}]}.

\bibitem{actpol}
S.~{Das}, B.~D. {Sherwin}, P.~{Aguirre}, J.~W. {Appel}, J.~R. {Bond}, C.~S.
  {Carvalho}, M.~J. {Devlin}, J.~{Dunkley}, R.~{D{\"u}nner},
  T.~{Essinger-Hileman}, J.~W. {Fowler}, A.~{Hajian}, M.~{Halpern},
  M.~{Hasselfield}, A.~D. {Hincks}, R.~{Hlozek}, K.~M. {Huffenberger}, J.~P.
  {Hughes}, K.~D. {Irwin}, J.~{Klein}, A.~{Kosowsky}, R.~H. {Lupton}, T.~A.
  {Marriage}, D.~{Marsden}, F.~{Menanteau}, K.~{Moodley}, M.~D. {Niemack},
  M.~R. {Nolta}, L.~A. {Page}, L.~{Parker}, E.~D. {Reese}, B.~L. {Schmitt},
  N.~{Sehgal}, J.~{Sievers}, D.~N. {Spergel}, S.~T. {Staggs}, D.~S. {Swetz},
  E.~R. {Switzer}, R.~{Thornton}, K.~{Visnjic}, and E.~{Wollack}.
\newblock {Detection of the Power Spectrum of Cosmic Microwave Background
  Lensing by the Atacama Cosmology Telescope}.
\newblock {\em Physical Review Letters}, 107(2):021301, July 2011.
\newblock \href {http://arxiv.org/abs/1103.2124} {\path{arXiv:1103.2124}},
  \href {http://dx.doi.org/10.1103/PhysRevLett.107.021301} {\path{[DOI]}},
  {\small[\href{http://adsabs.harvard.edu/abs/2011PhRvL.107b1301D}{ADS}]}.

\bibitem{Array:2016afx}
{BICEP2 Collaboration}, {Keck Array Collaboration}, P.~A.~R. {Ade}, Z.~{Ahmed},
  R.~W. {Aikin}, K.~D. {Alexander}, D.~{Barkats}, S.~J. {Benton}, C.~A.
  {Bischoff}, J.~J. {Bock}, R.~{Bowens-Rubin}, J.~A. {Brevik}, I.~{Buder},
  E.~{Bullock}, V.~{Buza}, J.~{Connors}, B.~P. {Crill}, L.~{Duband},
  C.~{Dvorkin}, J.~P. {Filippini}, S.~{Fliescher}, J.~{Grayson}, M.~{Halpern},
  S.~{Harrison}, S.~R. {Hildebrandt}, G.~C. {Hilton}, H.~{Hui}, K.~D. {Irwin},
  J.~{Kang}, K.~S. {Karkare}, E.~{Karpel}, J.~P. {Kaufman}, B.~G. {Keating},
  S.~{Kefeli}, S.~A. {Kernasovskiy}, J.~M. {Kovac}, C.~L. {Kuo}, E.~M.
  {Leitch}, M.~{Lueker}, K.~G. {Megerian}, T.~{Namikawa}, C.~B. {Netterfield},
  H.~T. {Nguyen}, R.~{O'Brient}, R.~W. {Ogburn}, IV, A.~{Orlando}, C.~{Pryke},
  S.~{Richter}, R.~{Schwarz}, C.~D. {Sheehy}, Z.~K. {Staniszewski},
  B.~{Steinbach}, R.~V. {Sudiwala}, G.~P. {Teply}, K.~L. {Thompson}, J.~E.
  {Tolan}, C.~{Tucker}, A.~D. {Turner}, A.~G. {Vieregg}, A.~C. {Weber}, D.~V.
  {Wiebe}, J.~{Willmert}, C.~L. {Wong}, W.~L.~K. {Wu}, and K.~W. {Yoon}.
\newblock {BICEP2/Keck Array VIII: Measurement of Gravitational Lensing from
  Large-scale B-mode Polarization}.
\newblock {\em \apj}, 833:228, December 2016.
\newblock \href {http://arxiv.org/abs/1606.01968} {\path{arXiv:1606.01968}},
  \href {http://dx.doi.org/10.3847/1538-4357/833/2/228} {\path{[DOI]}},
  {\small[\href{http://adsabs.harvard.edu/abs/2016ApJ...833..228B}{ADS}]}.

\bibitem{Ade:2013gez}
P.~A.~R. {Ade}, Y.~{Akiba}, A.~E. {Anthony}, K.~{Arnold}, M.~{Atlas},
  D.~{Barron}, D.~{Boettger}, J.~{Borrill}, S.~{Chapman}, Y.~{Chinone},
  M.~{Dobbs}, T.~{Elleflot}, J.~{Errard}, G.~{Fabbian}, C.~{Feng},
  D.~{Flanigan}, A.~{Gilbert}, W.~{Grainger}, N.~W. {Halverson}, M.~{Hasegawa},
  K.~{Hattori}, M.~{Hazumi}, W.~L. {Holzapfel}, Y.~{Hori}, J.~{Howard},
  P.~{Hyland}, Y.~{Inoue}, G.~C. {Jaehnig}, A.~{Jaffe}, B.~{Keating},
  Z.~{Kermish}, R.~{Keskitalo}, T.~{Kisner}, M.~{Le Jeune}, A.~T. {Lee},
  E.~{Linder}, E.~M. {Leitch}, M.~{Lungu}, F.~{Matsuda}, T.~{Matsumura},
  X.~{Meng}, N.~J. {Miller}, H.~{Morii}, S.~{Moyerman}, M.~J. {Myers},
  M.~{Navaroli}, H.~{Nishino}, H.~{Paar}, J.~{Peloton}, E.~{Quealy},
  G.~{Rebeiz}, C.~L. {Reichardt}, P.~L. {Richards}, C.~{Ross}, I.~{Schanning},
  D.~E. {Schenck}, B.~{Sherwin}, A.~{Shimizu}, C.~{Shimmin}, M.~{Shimon},
  P.~{Siritanasak}, G.~{Smecher}, H.~{Spieler}, N.~{Stebor}, B.~{Steinbach},
  R.~{Stompor}, A.~{Suzuki}, S.~{Takakura}, T.~{Tomaru}, B.~{Wilson},
  A.~{Yadav}, O.~{Zahn}, and {Polarbear Collaboration}.
\newblock {Measurement of the Cosmic Microwave Background Polarization Lensing
  Power Spectrum with the POLARBEAR Experiment}.
\newblock {\em Physical Review Letters}, 113(2):021301, July 2014.
\newblock \href {http://arxiv.org/abs/1312.6646} {\path{arXiv:1312.6646}},
  \href {http://dx.doi.org/10.1103/PhysRevLett.113.021301} {\path{[DOI]}},
  {\small[\href{http://adsabs.harvard.edu/abs/2014PhRvL.113b1301A}{ADS}]}.

\bibitem{boomerang}
C.~J. {MacTavish}, P.~A.~R. {Ade}, J.~J. {Bock}, J.~R. {Bond}, J.~{Borrill},
  A.~{Boscaleri}, P.~{Cabella}, C.~R. {Contaldi}, B.~P. {Crill}, P.~{de
  Bernardis}, G.~{De Gasperis}, A.~{de Oliveira-Costa}, G.~{De Troia}, G.~{di
  Stefano}, E.~{Hivon}, A.~H. {Jaffe}, W.~C. {Jones}, T.~S. {Kisner}, A.~E.
  {Lange}, A.~M. {Lewis}, S.~{Masi}, P.~D. {Mauskopf}, A.~{Melchiorri}, T.~E.
  {Montroy}, P.~{Natoli}, C.~B. {Netterfield}, E.~{Pascale}, F.~{Piacentini},
  D.~{Pogosyan}, G.~{Polenta}, S.~{Prunet}, S.~{Ricciardi}, G.~{Romeo}, J.~E.
  {Ruhl}, P.~{Santini}, M.~{Tegmark}, M.~{Veneziani}, and N.~{Vittorio}.
\newblock {Cosmological Parameters from the 2003 Flight of BOOMERANG}.
\newblock {\em \apj}, 647:799--812, August 2006.
\newblock \href {http://arxiv.org/abs/astro-ph/0507503}
  {\path{arXiv:astro-ph/0507503}}, \href {http://dx.doi.org/10.1086/505558}
  {\path{[DOI]}},
  {\small[\href{http://adsabs.harvard.edu/abs/2006ApJ...647..799M}{ADS}]}.

\bibitem{Stompor:2003kr}
R.~{Stompor}, S.~{Hanany}, M.~E. {Abroe}, J.~{Borrill}, P.~G. {Ferreira}, A.~H.
  {Jaffe}, B.~{Johnson}, A.~T. {Lee}, B.~{Rabii}, P.~L. {Richards}, G.~{Smoot},
  C.~{Winant}, and J.~H.~P. {Wu}.
\newblock {The MAXIMA Experiment: Latest Results and Consistency Tests}.
\newblock {\em ArXiv Astrophysics e-prints}, September 2003.
\newblock \href {http://arxiv.org/abs/astro-ph/0309409}
  {\path{arXiv:astro-ph/0309409}},
  {\small[\href{http://adsabs.harvard.edu/abs/2003astro.ph..9409S}{ADS}]}.

\bibitem{2012sngi.confE...3M}
Y.~{Mellier}.
\newblock {Euclid: Mapping the Geometry of the Dark Universe}.
\newblock In {\em Science from the Next Generation Imaging and Spectroscopic
  Surveys}, page~3, October 2012.
\newblock
  {\small[\href{http://adsabs.harvard.edu/abs/2012sngi.confE...3M}{ADS}]}.

\bibitem{2009arXiv0912.0201L}
{LSST Science Collaboration}, P.~A. {Abell}, J.~{Allison}, S.~F. {Anderson},
  J.~R. {Andrew}, J.~R.~P. {Angel}, L.~{Armus}, D.~{Arnett}, S.~J. {Asztalos},
  T.~S. {Axelrod}, and et~al.
\newblock {LSST Science Book, Version 2.0}.
\newblock {\em ArXiv e-prints}, December 2009.
\newblock \href {http://arxiv.org/abs/0912.0201} {\path{arXiv:0912.0201}},
  {\small[\href{http://adsabs.harvard.edu/abs/2009arXiv0912.0201L}{ADS}]}.

\bibitem{Carilli:2004nx}
Chris~L. Carilli and S.~Rawlings.
\newblock {Science with the Square Kilometer Array: Motivation, key science
  projects, standards and assumptions}.
\newblock {\em New Astron. Rev.}, 48:979, 2004.
\newblock \href {http://arxiv.org/abs/astro-ph/0409274}
  {\path{arXiv:astro-ph/0409274}}, \href
  {http://dx.doi.org/10.1016/j.newar.2004.09.001} {\path{[DOI]}}.

\bibitem{Kogut:2011xw}
A.~{Kogut}, D.~J. {Fixsen}, D.~T. {Chuss}, J.~{Dotson}, E.~{Dwek},
  M.~{Halpern}, G.~F. {Hinshaw}, S.~M. {Meyer}, S.~H. {Moseley}, M.~D.
  {Seiffert}, D.~N. {Spergel}, and E.~J. {Wollack}.
\newblock {The Primordial Inflation Explorer (PIXIE): a nulling polarimeter for
  cosmic microwave background observations}.
\newblock {\em \jcap}, 7:25, 2011.
\newblock \href {http://arxiv.org/abs/1105.2044} {\path{arXiv:1105.2044}},
  \href {http://dx.doi.org/10.1088/1475-7516/2011/07/025} {\path{[DOI]}},
  {\small[\href{http://adsabs.harvard.edu/abs/2011JCAP...07..025K}{ADS}]}.

\bibitem{Matsumura:2013aja}
T.~Matsumura et~al.
\newblock {Mission design of LiteBIRD}.
\newblock 2013.
\newblock [J. Low. Temp. Phys.176,733(2014)].
\newblock \href {http://arxiv.org/abs/1311.2847} {\path{arXiv:1311.2847}},
  \href {http://dx.doi.org/10.1007/s10909-013-0996-1} {\path{[DOI]}}.

\bibitem{Lewis:2002ah}
Antony Lewis and Sarah Bridle.
\newblock {Cosmological parameters from CMB and other data: A Monte Carlo
  approach}.
\newblock {\em Phys. Rev.}, D66:103511, 2002.
\newblock \href {http://arxiv.org/abs/astro-ph/0205436}
  {\path{arXiv:astro-ph/0205436}}, \href
  {http://dx.doi.org/10.1103/PhysRevD.66.103511} {\path{[DOI]}}.

\bibitem{Javanmardi:2015sfa}
Behnam Javanmardi, Cristiano Porciani, Pavel Kroupa, and Jan Pflamm-Altenburg.
\newblock {Probing the isotropy of cosmic acceleration traced by Type Ia
  supernovae}.
\newblock {\em Astrophys. J.}, 810(1):47, 2015.
\newblock \href {http://arxiv.org/abs/1507.07560} {\path{arXiv:1507.07560}},
  \href {http://dx.doi.org/10.1088/0004-637X/810/1/47} {\path{[DOI]}}.

\bibitem{Carvalho:2015lqd}
C.~Sofia Carvalho and Katrine Marques.
\newblock {Angular distribution of cosmological parameters as a probe of
  space-time inhomogeneities}.
\newblock {\em Astron. Astrophys.}, 592:A102, 2016.
\newblock \href {http://arxiv.org/abs/1512.07869} {\path{arXiv:1512.07869}},
  \href {http://dx.doi.org/10.1051/0004-6361/201628177} {\path{[DOI]}}.

\bibitem{Carvalho:2016aah}
C.~Sofia Carvalho and Spyros Basilakos.
\newblock {Angular distribution of cosmological parameters as a probe of
  inhomogeneities: a kinematic parametrisation}.
\newblock {\em Astron. Astrophys.}, 592:A152, 2016.
\newblock \href {http://arxiv.org/abs/1603.07519} {\path{arXiv:1603.07519}},
  \href {http://dx.doi.org/10.1051/0004-6361/201628572} {\path{[DOI]}}.

\bibitem{Cai:2011xs}
Rong-Gen Cai and Zhong-Liang Tuo.
\newblock {Direction Dependence of the Deceleration Parameter}.
\newblock {\em JCAP}, 1202:004, 2012.
\newblock \href {http://arxiv.org/abs/1109.0941} {\path{arXiv:1109.0941}},
  \href {http://dx.doi.org/10.1088/1475-7516/2012/02/004} {\path{[DOI]}}.

\bibitem{2013ApJ...773L...3A}
M.~{Axelsson}, Y.~{Fantaye}, F.~K. {Hansen}, A.~J. {Banday}, H.~K. {Eriksen},
  and K.~M. {Gorski}.
\newblock {Directional Dependence of {$\Lambda$}CDM Cosmological Parameters}.
\newblock {\em \apjl}, 773:L3, August 2013.
\newblock \href {http://arxiv.org/abs/1303.5371} {\path{arXiv:1303.5371}},
  \href {http://dx.doi.org/10.1088/2041-8205/773/1/L3} {\path{[DOI]}},
  {\small[\href{http://adsabs.harvard.edu/abs/2013ApJ...773L...3A}{ADS}]}.

\bibitem{2016JCAP...06..042M}
S.~{Mukherjee}, P.~K. {Aluri}, S.~{Das}, S.~{Shaikh}, and T.~{Souradeep}.
\newblock {Direction dependence of cosmological parameters due to cosmic
  hemispherical asymmetry}.
\newblock {\em \jcap}, 6:042, June 2016.
\newblock \href {http://arxiv.org/abs/1510.00154} {\path{arXiv:1510.00154}},
  \href {http://dx.doi.org/10.1088/1475-7516/2016/06/042} {\path{[DOI]}},
  {\small[\href{http://adsabs.harvard.edu/abs/2016JCAP...06..042M}{ADS}]}.

\bibitem{2015JCAP...11..012H}
D.~K. {Hazra} and A.~{Shafieloo}.
\newblock {Search for a direction in the forest of Lyman-{$\alpha$}}.
\newblock {\em \jcap}, 11:012, November 2015.
\newblock \href {http://arxiv.org/abs/1506.03926} {\path{arXiv:1506.03926}},
  \href {http://dx.doi.org/10.1088/1475-7516/2015/11/012} {\path{[DOI]}},
  {\small[\href{http://adsabs.harvard.edu/abs/2015JCAP...11..012H}{ADS}]}.

\bibitem{2001PThPh.105..419T}
K.~{Tomita}.
\newblock {Anisotropy of the Hubble Constant in a Cosmological Model with a
  Local Void on Scales of \~{} 200 Mpc}.
\newblock {\em Progress of Theoretical Physics}, 105:419--427, March 2001.
\newblock \href {http://arxiv.org/abs/astro-ph/0005031}
  {\path{arXiv:astro-ph/0005031}}, \href
  {http://dx.doi.org/10.1143/PTP.105.419} {\path{[DOI]}},
  {\small[\href{http://adsabs.harvard.edu/abs/2001PThPh.105..419T}{ADS}]}.

\bibitem{McClure2007533}
M.L. McClure and C.C. Dyer.
\newblock Anisotropy in the hubble constant as observed in the \{HST\}
  extragalactic distance scale key project results.
\newblock {\em New Astronomy}, 12(7):533 -- 543, 2007.
\newblock URL:
  \url{http://www.sciencedirect.com/science/article/pii/S1384107607000231},
  \href {http://dx.doi.org/https://doi.org/10.1016/j.newast.2007.03.005}
  {\path{[DOI]}}.

\bibitem{Romano:2014iea}
Antonio~Enea Romano and Sergio~Andrés Vallejo.
\newblock {Directional dependence of the local estimation of $H_0$ and the
  nonperturbative effects of primordial curvature perturbations}.
\newblock {\em Europhys. Lett.}, 109(3):39002, 2015.
\newblock \href {http://arxiv.org/abs/1403.2034} {\path{arXiv:1403.2034}},
  \href {http://dx.doi.org/10.1209/0295-5075/109/39002} {\path{[DOI]}}.

\bibitem{PhysRevD.93.083525}
Alan Zablocki and Scott Dodelson.
\newblock Extreme data compression for the cmb.
\newblock {\em Phys. Rev. D}, 93:083525, Apr 2016.
\newblock URL: \url{https://link.aps.org/doi/10.1103/PhysRevD.93.083525}, \href
  {http://dx.doi.org/10.1103/PhysRevD.93.083525} {\path{[DOI]}}.

\bibitem{Mukherjee:2015wra}
Suvodip Mukherjee and Tarun Souradeep.
\newblock {Litmus Test for Cosmic Hemispherical Asymmetry in the Cosmic
  Microwave Background B-mode polarization}.
\newblock {\em Phys. Rev. Lett.}, 116(22):221301, 2016.
\newblock \href {http://arxiv.org/abs/1509.06736} {\path{arXiv:1509.06736}},
  \href {http://dx.doi.org/10.1103/PhysRevLett.116.221301} {\path{[DOI]}}.

\bibitem{1998CQGra..15.2657C}
N.~J. {Cornish}, D.~N. {Spergel}, and G.~D. {Starkman}.
\newblock {Circles in the sky: finding topology with the microwave background
  radiation}.
\newblock {\em Classical and Quantum Gravity}, 15:2657--2670, September 1998.
\newblock \href {http://arxiv.org/abs/gr-qc/9602039}
  {\path{arXiv:gr-qc/9602039}}, \href
  {http://dx.doi.org/10.1088/0264-9381/15/9/013} {\path{[DOI]}},
  {\small[\href{http://adsabs.harvard.edu/abs/1998CQGra..15.2657C}{ADS}]}.

\bibitem{2011ApJ...740...52H}
A.~{Hajian}.
\newblock {Are there Echoes from the Pre-big-bang Universe? A Search for
  Low-variance Circles in the Cosmic Microwave Background Sky}.
\newblock {\em \apj}, 740:52, October 2011.
\newblock \href {http://arxiv.org/abs/1012.1656} {\path{arXiv:1012.1656}},
  \href {http://dx.doi.org/10.1088/0004-637X/740/2/52} {\path{[DOI]}},
  {\small[\href{http://adsabs.harvard.edu/abs/2011ApJ...740...52H}{ADS}]}.

\bibitem{2010ApJ...724..374K}
E.~D. {Kovetz}, A.~{Ben-David}, and N.~{Itzhaki}.
\newblock {Giant Rings in the Cosmic Microwave Background Sky}.
\newblock {\em \apj}, 724:374--378, November 2010.
\newblock \href {http://arxiv.org/abs/1005.3923} {\path{arXiv:1005.3923}},
  \href {http://dx.doi.org/10.1088/0004-637X/724/1/374} {\path{[DOI]}},
  {\small[\href{http://adsabs.harvard.edu/abs/2010ApJ...724..374K}{ADS}]}.

\bibitem{2011JCAP...04..033M}
A.~{Moss}, D.~{Scott}, and J.~P. {Zibin}.
\newblock {No evidence for anomalously low variance circles on the sky}.
\newblock {\em \jcap}, 4:033, April 2011.
\newblock \href {http://arxiv.org/abs/1012.1305} {\path{arXiv:1012.1305}},
  \href {http://dx.doi.org/10.1088/1475-7516/2011/04/033} {\path{[DOI]}},
  {\small[\href{http://adsabs.harvard.edu/abs/2011JCAP...04..033M}{ADS}]}.

\bibitem{2013MNRAS.429.1376B}
P.~{Bielewicz}, B.~D. {Wandelt}, and A.~J. {Banday}.
\newblock {A search for concentric rings with unusual variance in the 7-year
  WMAP temperature maps using a fast convolution approach}.
\newblock {\em \mnras}, 429:1376--1385, February 2013.
\newblock \href {http://arxiv.org/abs/1207.6905} {\path{arXiv:1207.6905}},
  \href {http://dx.doi.org/10.1093/mnras/sts424} {\path{[DOI]}},
  {\small[\href{http://adsabs.harvard.edu/abs/2013MNRAS.429.1376B}{ADS}]}.

\bibitem{2010JCAP...02..004F}
A.~{Fialkov}, N.~{Itzhaki}, and E.~D. {Kovetz}.
\newblock {Cosmological imprints of pre-inflationary particles}.
\newblock {\em \jcap}, 2:004, February 2010.
\newblock \href {http://arxiv.org/abs/0911.2100} {\path{arXiv:0911.2100}},
  \href {http://dx.doi.org/10.1088/1475-7516/2010/02/004} {\path{[DOI]}},
  {\small[\href{http://adsabs.harvard.edu/abs/2010JCAP...02..004F}{ADS}]}.

\bibitem{2011PhRvD..84d3507F}
S.~M. {Feeney}, M.~C. {Johnson}, D.~J. {Mortlock}, and H.~V. {Peiris}.
\newblock {First observational tests of eternal inflation: Analysis methods and
  WMAP 7-year results}.
\newblock {\em \prd}, 84(4):043507, August 2011.
\newblock \href {http://arxiv.org/abs/1012.3667} {\path{arXiv:1012.3667}},
  \href {http://dx.doi.org/10.1103/PhysRevD.84.043507} {\path{[DOI]}},
  {\small[\href{http://adsabs.harvard.edu/abs/2011PhRvD..84d3507F}{ADS}]}.

\bibitem{Tegmark:1996qt}
Max Tegmark.
\newblock {How to measure CMB power spectra without losing information}.
\newblock {\em Phys. Rev.}, D55:5895--5907, 1997.
\newblock \href {http://arxiv.org/abs/astro-ph/9611174}
  {\path{arXiv:astro-ph/9611174}}, \href
  {http://dx.doi.org/10.1103/PhysRevD.55.5895} {\path{[DOI]}}.

\bibitem{Bond:1998zw}
J.~R. Bond, Andrew~H. Jaffe, and L.~Knox.
\newblock {Estimating the power spectrum of the cosmic microwave background}.
\newblock {\em Phys. Rev.}, D57:2117--2137, 1998.
\newblock \href {http://arxiv.org/abs/astro-ph/9708203}
  {\path{arXiv:astro-ph/9708203}}, \href
  {http://dx.doi.org/10.1103/PhysRevD.57.2117} {\path{[DOI]}}.

\bibitem{Oh:1998sr}
Siang~Peng Oh, David~N. Spergel, and Gary Hinshaw.
\newblock {An Efficient technique to determine the power spectrum from cosmic
  microwave background sky maps}.
\newblock {\em Astrophys. J.}, 510:551, 1999.
\newblock \href {http://arxiv.org/abs/astro-ph/9805339}
  {\path{arXiv:astro-ph/9805339}}, \href {http://dx.doi.org/10.1086/306629}
  {\path{[DOI]}}.

\bibitem{Wandelt:2003uk}
Benjamin~D. Wandelt, David~L. Larson, and Arun Lakshminarayanan.
\newblock {Global, exact cosmic microwave background data analysis using Gibbs
  sampling}.
\newblock {\em Phys. Rev.}, D70:083511, 2004.
\newblock \href {http://arxiv.org/abs/astro-ph/0310080}
  {\path{arXiv:astro-ph/0310080}}, \href
  {http://dx.doi.org/10.1103/PhysRevD.70.083511} {\path{[DOI]}}.

\bibitem{Elsner:2012kj}
Franz Elsner and Benjamin~D. Wandelt.
\newblock {Fast calculation of the Fisher matrix for Cosmic Microwave
  Background experiments}.
\newblock {\em Astron. Astrophys.}, 540:L6, 2012.
\newblock \href {http://arxiv.org/abs/1202.4898} {\path{arXiv:1202.4898}},
  \href {http://dx.doi.org/10.1051/0004-6361/201218985} {\path{[DOI]}}.

\bibitem{Hinshaw:2003fc}
G.~Hinshaw et~al.
\newblock {First year Wilkinson Microwave Anisotropy Probe (WMAP) observations:
  Data processing methods and systematic errors limits}.
\newblock {\em Astrophys. J. Suppl.}, 148:63, 2003.
\newblock \href {http://arxiv.org/abs/astro-ph/0302222}
  {\path{arXiv:astro-ph/0302222}}, \href {http://dx.doi.org/10.1086/377222}
  {\path{[DOI]}}.

\bibitem{Dunkley:2008ie}
J.~Dunkley et~al.
\newblock {Five-Year Wilkinson Microwave Anisotropy Probe (WMAP) Observations:
  Likelihoods and Parameters from the WMAP data}.
\newblock {\em Astrophys. J. Suppl.}, 180:306--329, 2009.
\newblock \href {http://arxiv.org/abs/0803.0586} {\path{arXiv:0803.0586}},
  \href {http://dx.doi.org/10.1088/0067-0049/180/2/306} {\path{[DOI]}}.

\bibitem{Elsner:2012bq}
Franz Elsner and Benjamin~D. Wandelt.
\newblock {Likelihood, Fisher information, and systematics of cosmic microwave
  background experiments}.
\newblock {\em Astron. Astrophys.}, 542:A60, 2012.
\newblock \href {http://arxiv.org/abs/1205.0810} {\path{arXiv:1205.0810}},
  \href {http://dx.doi.org/10.1051/0004-6361/201219293} {\path{[DOI]}}.

\bibitem{Tegmark:1996qs}
Max Tegmark.
\newblock {How to make maps from CMB data without losing information}.
\newblock {\em Astrophys. J.}, 480:L87--L90, 1997.
\newblock \href {http://arxiv.org/abs/astro-ph/9611130}
  {\path{arXiv:astro-ph/9611130}}, \href {http://dx.doi.org/10.1086/310631}
  {\path{[DOI]}}.

\bibitem{Hirata:2004rp}
Christopher~M. Hirata, Nikhil Padmanabhan, Uros Seljak, David Schlegel, and
  Jonathan Brinkmann.
\newblock {Cross-correlation of CMB with large-scale structure: Weak
  gravitational lensing}.
\newblock {\em Phys. Rev.}, D70:103501, 2004.
\newblock \href {http://arxiv.org/abs/astro-ph/0406004}
  {\path{arXiv:astro-ph/0406004}}, \href
  {http://dx.doi.org/10.1103/PhysRevD.70.103501} {\path{[DOI]}}.

\bibitem{Smith:2007rg}
Kendrick~M. Smith, Oliver Zahn, and Olivier Dore.
\newblock {Detection of Gravitational Lensing in the Cosmic Microwave
  Background}.
\newblock {\em Phys. Rev.}, D76:043510, 2007.
\newblock \href {http://arxiv.org/abs/0705.3980} {\path{arXiv:0705.3980}},
  \href {http://dx.doi.org/10.1103/PhysRevD.76.043510} {\path{[DOI]}}.

\bibitem{Elsner:2012fe}
Franz Elsner and Benjamin~D. Wandelt.
\newblock {Efficient Wiener filtering without preconditioning}.
\newblock {\em Astron. Astrophys.}, 549:A111, 2013.
\newblock \href {http://arxiv.org/abs/1210.4931} {\path{arXiv:1210.4931}},
  \href {http://dx.doi.org/10.1051/0004-6361/201220586} {\path{[DOI]}}.

\bibitem{Bunn:2016lxi}
Emory~F. Bunn and Benjamin Wandelt.
\newblock {Pure E and B polarization maps via Wiener filtering}.
\newblock 2016.
\newblock \href {http://arxiv.org/abs/1610.03345} {\path{arXiv:1610.03345}}.

\bibitem{Ramanah:2017luo}
Doogesh~Kodi Ramanah, Guilhem Lavaux, and Benjamin~D. Wandelt.
\newblock {Wiener filter reloaded: fast signal reconstruction without
  preconditioning}.
\newblock {\em Mon. Not. Roy. Astron. Soc.}, 468(2):1782--1793, 2017.
\newblock \href {http://arxiv.org/abs/1702.08852} {\path{arXiv:1702.08852}},
  \href {http://dx.doi.org/10.1093/mnras/stx527} {\path{[DOI]}}.

\bibitem{2002MNRAS.334..167G}
S.~{Gupta} and A.~F. {Heavens}.
\newblock {Fast parameter estimation from the cosmic microwave background power
  spectrum}.
\newblock {\em \mnras}, 334:167--172, July 2002.
\newblock \href {http://arxiv.org/abs/astro-ph/0108315}
  {\path{arXiv:astro-ph/0108315}}, \href
  {http://dx.doi.org/10.1046/j.1365-8711.2002.05499.x} {\path{[DOI]}},
  {\small[\href{http://adsabs.harvard.edu/abs/2002MNRAS.334..167G}{ADS}]}.

\bibitem{smica}
Planck~2015 results: SMICA Temperature~Map.

\bibitem{1973ApJ...185..757H}
M.~G. {Hauser} and P.~J.~E. {Peebles}.
\newblock {Statistical Analysis of Catalogs of Extragalactic Objects. II. the
  Abell Catalog of Rich Clusters}.
\newblock {\em \apj}, 185:757--786, November 1973.
\newblock \href {http://dx.doi.org/10.1086/152453} {\path{[DOI]}},
  {\small[\href{http://adsabs.harvard.edu/abs/1973ApJ...185..757H}{ADS}]}.

\bibitem{Wandelt:2000av}
Benjamin~D. Wandelt, Eric Hivon, and Krzysztof~M. Gorski.
\newblock {The pseudo-$c_l$ method: cosmic microwave background anisotropy
  power spectrum statistics for high precision cosmology}.
\newblock {\em Phys. Rev.}, D64:083003, 2001.
\newblock \href {http://arxiv.org/abs/astro-ph/0008111}
  {\path{arXiv:astro-ph/0008111}}, \href
  {http://dx.doi.org/10.1103/PhysRevD.64.083003} {\path{[DOI]}}.

\bibitem{Hivon:2001jp}
E.~Hivon, K.~M. Gorski, C.~B. Netterfield, B.~P. Crill, S.~Prunet, and
  F.~Hansen.
\newblock {Master of the cosmic microwave background anisotropy power spectrum:
  a fast method for statistical analysis of large and complex cosmic microwave
  background data sets}.
\newblock {\em Astrophys. J.}, 567:2, 2002.
\newblock \href {http://arxiv.org/abs/astro-ph/0105302}
  {\path{arXiv:astro-ph/0105302}}, \href {http://dx.doi.org/10.1086/338126}
  {\path{[DOI]}}.

\bibitem{Tegmark:1996bz}
Max Tegmark, Andy Taylor, and Alan Heavens.
\newblock {Karhunen-Loeve eigenvalue problems in cosmology: How should we
  tackle large data sets?}
\newblock {\em Astrophys. J.}, 480:22, 1997.
\newblock \href {http://arxiv.org/abs/astro-ph/9603021}
  {\path{arXiv:astro-ph/9603021}}, \href {http://dx.doi.org/10.1086/303939}
  {\path{[DOI]}}.

\bibitem{Gorski:2004by}
K.~M. Gorski, Eric Hivon, A.~J. Banday, B.~D. Wandelt, F.~K. Hansen,
  M.~Reinecke, and M.~Bartelman.
\newblock {HEALPix - A Framework for high resolution discretization, and fast
  analysis of data distributed on the sphere}.
\newblock {\em Astrophys. J.}, 622:759--771, 2005.
\newblock \href {http://arxiv.org/abs/astro-ph/0409513}
  {\path{arXiv:astro-ph/0409513}}, \href {http://dx.doi.org/10.1086/427976}
  {\path{[DOI]}}.

\bibitem{Ade:2015xua}
P.~A.~R. Ade et~al.
\newblock {Planck 2015 results. XIII. Cosmological parameters}.
\newblock {\em Astron. Astrophys.}, 594:A13, 2016.
\newblock \href {http://arxiv.org/abs/1502.01589} {\path{arXiv:1502.01589}},
  \href {http://dx.doi.org/10.1051/0004-6361/201525830} {\path{[DOI]}}.

\bibitem{Lewis:1999bs}
Antony Lewis, Anthony Challinor, and Anthony Lasenby.
\newblock {Efficient computation of CMB anisotropies in closed FRW models}.
\newblock {\em Astrophys. J.}, 538:473--476, 2000.
\newblock \href {http://arxiv.org/abs/astro-ph/9911177}
  {\path{arXiv:astro-ph/9911177}}, \href {http://dx.doi.org/10.1086/309179}
  {\path{[DOI]}}.

\bibitem{Aghanim:2016fhp}
N.~Aghanim et~al.
\newblock {Planck intermediate results. XLIX. Parity-violation constraints from
  polarization data}.
\newblock {\em Astron. Astrophys.}, 596:A110, 2016.
\newblock \href {http://arxiv.org/abs/1605.08633} {\path{arXiv:1605.08633}},
  \href {http://dx.doi.org/10.1051/0004-6361/201629018} {\path{[DOI]}}.

\end{thebibliography}
\end{document}